\documentclass[final]{cvpr}

\usepackage{times}
\usepackage{epsfig}
\usepackage{graphicx}
\usepackage{amsmath}
\usepackage{amssymb}
\usepackage{graphicx}
\usepackage{multirow}

\usepackage{amsmath,amssymb}

\usepackage{color}
\usepackage{cite}
\usepackage{dsfont}
\usepackage[table,xcdraw]{xcolor}
\usepackage{ragged2e}
\usepackage{nccmath}
\usepackage{textcomp}
\usepackage{arydshln}
\usepackage{amsmath}
\usepackage{caption}

\usepackage[pagebackref=true,breaklinks=true,colorlinks,bookmarks=false]{hyperref}

\def\cvprPaperID{1532} 
\def\confYear{CVPR 2021}
\begin{document}

\title{Learning Multi-Scale Photo Exposure Correction}

\author{Mahmoud Afifi$^{1,2}$\thanks{This work was done while Mahmoud Afifi was an intern at the SAIC.} \hspace{5mm} Konstantinos G. Derpanis$^{1}$ \hspace{5mm} Bj{\"o}rn Ommer$^{3}$ \hspace{5mm} Michael S. Brown$^{1}$\\\\
$^{1}$Samsung AI Centre (SAIC), Toronto, Canada\\$^{2}$York University, Canada\hspace{0.4cm}$^{3}$Heidelberg University, Germany}

\maketitle

\begin{abstract}
Capturing photographs with wrong exposures remains a major source of errors in camera-based imaging. Exposure problems are categorized as either: (i) overexposed, where the camera exposure was too long, resulting in bright and washed-out image regions, or (ii) underexposed, where the exposure was too short, resulting in dark regions. Both under- and overexposure greatly reduce the contrast and visual appeal of an image. Prior work mainly focuses on underexposed images or general image enhancement. In contrast, our proposed method targets both over- and underexposure errors in photographs. We formulate the exposure correction problem as two main sub-problems: (i) color enhancement and (ii) detail enhancement. Accordingly, we propose a coarse-to-fine deep neural network (DNN) model, trainable in an end-to-end manner, that addresses each sub-problem separately. A key aspect of our solution is a new dataset of over 24,000 images exhibiting the broadest range of exposure values to date with a corresponding properly exposed image. Our method achieves results on par with existing state-of-the-art methods on underexposed images and yields significant improvements for images suffering from overexposure errors.
\end{abstract}

\section{Introduction}
\label{sec:intro}

The exposure used at capture time directly affects the overall brightness of the final rendered photograph. Digital cameras control exposure using three main factors: (i) capture shutter speed, (ii) f-number, which is the ratio of the focal length to the camera aperture diameter, and (iii) the ISO value to control the amplification factor of the received pixel signals. In photography, exposure settings are represented by exposure values (EVs), where each EV refers to different combinations of camera shutter speeds and f-numbers that result in the same exposure effect---also referred to as `equivalent exposures' in photography.

\begin{figure}[t]
\centering
\includegraphics[width=\linewidth]{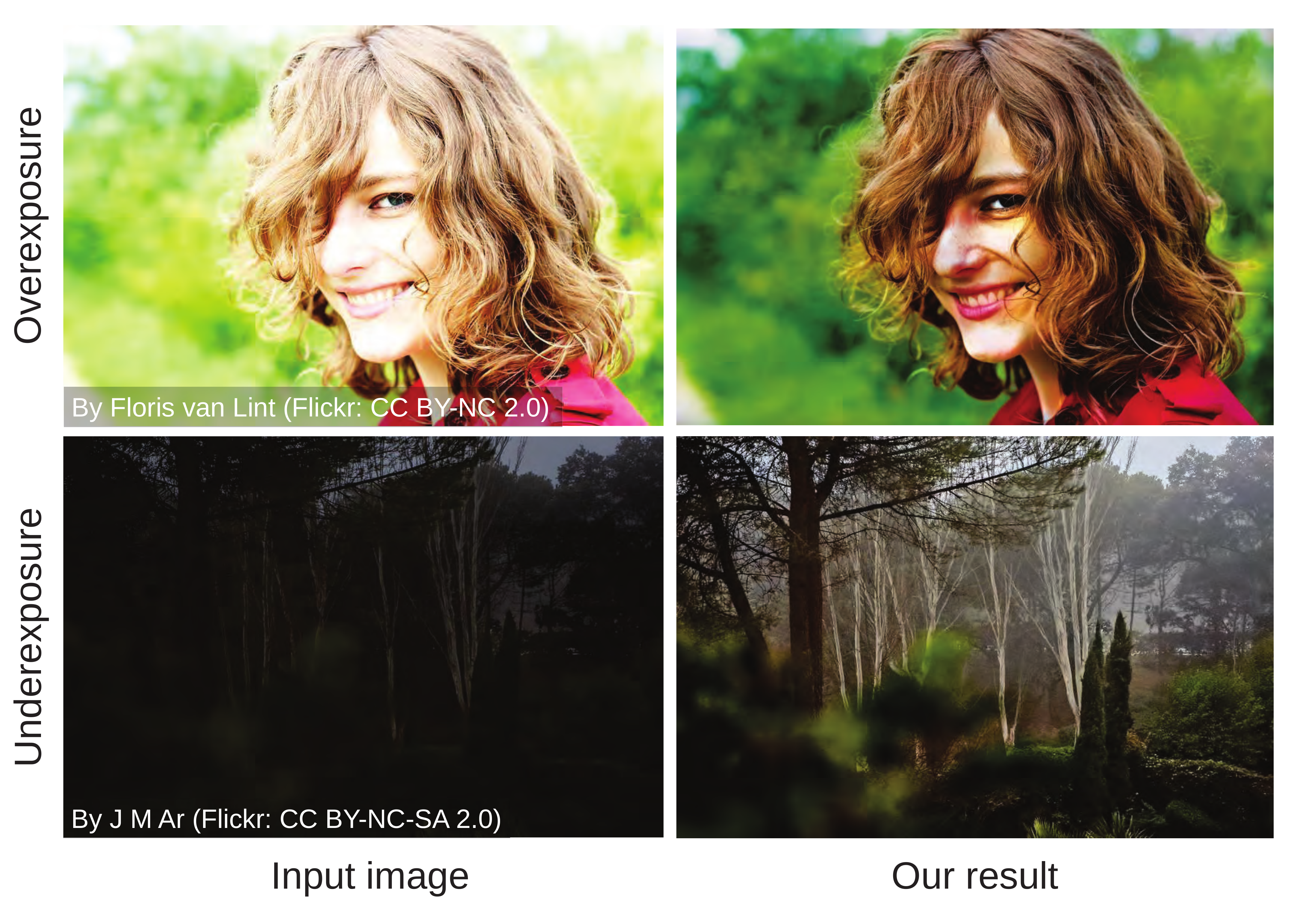}
\vspace{-6mm}
\caption{Photographs with over- and underexposure errors and the results of our method using a \textit{single} model for exposure correction. These sample input images are taken from outside our dataset to demonstrate the generalization of our trained model.\vspace{-4mm}}
\label{fig:teaser}
\end{figure}

Digital cameras can adjust the exposure value of captured images for the purpose of varying the brightness levels. This adjustment can be controlled manually by users or performed automatically in an auto-exposure (AE) mode. When AE is used, cameras adjust the EV to compensate for low/high levels of brightness in the captured scene using through-the-lens (TTL) metering that measures the amount of light received from the scene \cite{peterson2016understanding}.

Exposure errors can occur due to several factors, such as errors in measurements of TTL metering, hard lighting conditions (e.g., very low lighting and backlighting), dramatic changes in the brightness level of the scene, and errors made by users in the manual mode. Such exposure errors are introduced early in the capture process and are thus hard to correct after rendering the final 8-bit image. This is due to the highly nonlinear operations applied by the camera image signal processor (ISP) afterwards to render the final 8-bit standard RGB (sRGB) image \cite{karaimer2016software}.

Fig.\ \ref{fig:teaser} shows typical examples of images with exposure errors. In Fig.\ \ref{fig:teaser}, exposure errors result in either very bright image regions, due to overexposure, or very dark regions, caused by underexposure errors, in the final rendered images. Correcting images with such errors is a challenging task even for well-established image enhancement software packages, see Fig.\ \ref{fig:ours_vs_commercial_sw_supp}. Although both over- and underexposure errors are common in photography, most prior work is mainly focused on correcting underexposure errors \cite{guo2017lime, HQEC, Chen2018Retinex, zhang2019kindling, DeepUPE} or generic image quality enhancement \cite{HDRNET, DPE}.

\noindent\textbf{Contributions }
We propose a coarse-to-fine deep learning method for exposure error correction of \textit{both} over- and underexposed sRGB images. Our approach formulates the exposure correction problem as two main sub-problems: (i) color and (ii) detail enhancement. We propose a coarse-to-fine deep
neural network (DNN) model, trainable in an end-to-end manner, that
begins by correcting the global color information and subsequently refines the image details.
In addition to our DNN model, a key contribution to the exposure correction problem is a new dataset containing over 24,000 images\footnote{Project page: \href{https://github.com/mahmoudnafifi/Exposure_Correction}{https://github.com/mahmoudnafifi/Exposure\_Correction}} rendered from raw-RGB to sRGB with different exposure settings with broader exposure ranges than previous datasets. Each image in our dataset is provided with a corresponding properly exposed reference image. Lastly, we present an extensive set of evaluations and ablations of our proposed method with comparisons to the state of the art.  We demonstrate that our method achieves results on par with previous methods dedicated to underexposed images and yields significant improvements on overexposed images. Furthermore, our model generalizes well to images outside our dataset.

\section{Related Work}
\label{sec:relatedwork}
The focus of our paper is on correcting exposure errors in camera-rendered 8-bit sRGB images. We refer the reader to \cite{chen2018learning, hu2018exposure, hasinoff2016burst, liba2019handheld} for representative examples for rendering linear raw-RGB images captured with low light or exposure errors.

\noindent\textbf{Exposure Correction }  Traditional methods for exposure correction and contrast enhancement rely on image histograms to re-balance image intensity values \cite{10.5555/559707, pizer1987adaptive, adaptivehisteq, 5773086, lee2013contrast}.  Alternatively,
tone curve adjustment is used to correct images with exposure errors. This process is performed by relying either solely on input image information \cite{yuan2012automatic} or trained deep learning models \cite{yu2018deepexposure, park2018distort, guo2020zero, moran2020deeplpf}.
The majority of prior work adopts the Retinex theory \cite{land1977retinex} by assuming that improperly exposed images can be formulated as a pixel-wise multiplication of target images, captured with correct exposure settings, by illumination maps. Thus, the goal of these methods is to predict illumination maps to recover the well-exposed target images. Representative Retinex-based methods include \cite{land1977retinex, jobson1997multiscale, wang2013naturalness, meylan2006high, guo2017lime, HQEC, zhang2019dual} and the most recent deep learning ones \cite{Chen2018Retinex, zhang2019kindling, DeepUPE}. Most of these methods, however, are restricted to correcting underexposure errors \cite{guo2017lime, HQEC, Chen2018Retinex, zhang2019kindling, DeepUPE, yang2020fidelity, xu2020learning, zhu2020eemefn}.
In contrast to the majority of prior work, our work is the first deep learning method to explicitly correct {\it both} overexposed and underexposed photographs with a single model.

\noindent\textbf{HDR Restoration and Image Enhancement } HDR restoration is the process of reconstructing scene radiance HDR values from one or more low dynamic range (LDR) input images. Prior work either require access to multiple LDR images \cite{mertens2009exposure, kalantari2017deep, endoSA2017} or use a single LDR input image, which is converted to an HDR image by hallucinating missing information~\cite{HDRCNN, moriwaki2018hybrid}.
Ultimately, these reconstructed HDR images are mapped back to LDR for perceptual visualization. This mapping can be directly performed from the input multi-LDR images \cite{debevec1997recovering,cai2018learning}, the reconstructed HDR image \cite{yang2018image}, or directly from the single input LDR image without the need for radiance HDR reconstruction \cite{HDRNET, DPE}.   There are also methods that focus on general image enhancement that can be applied to enhancing images with poor exposure.  In particular, work by~\cite{DPED, WESPE} was developed primarily to enhance images captured on smartphone cameras by mapping captured images to appear as high-quality images captured by a DSLR.  Our work does not seek to reconstruct HDR images or general enhancement, but instead is trained to explicitly address exposure errors.

\begin{figure}[t]
\centering
\includegraphics[width=\linewidth]{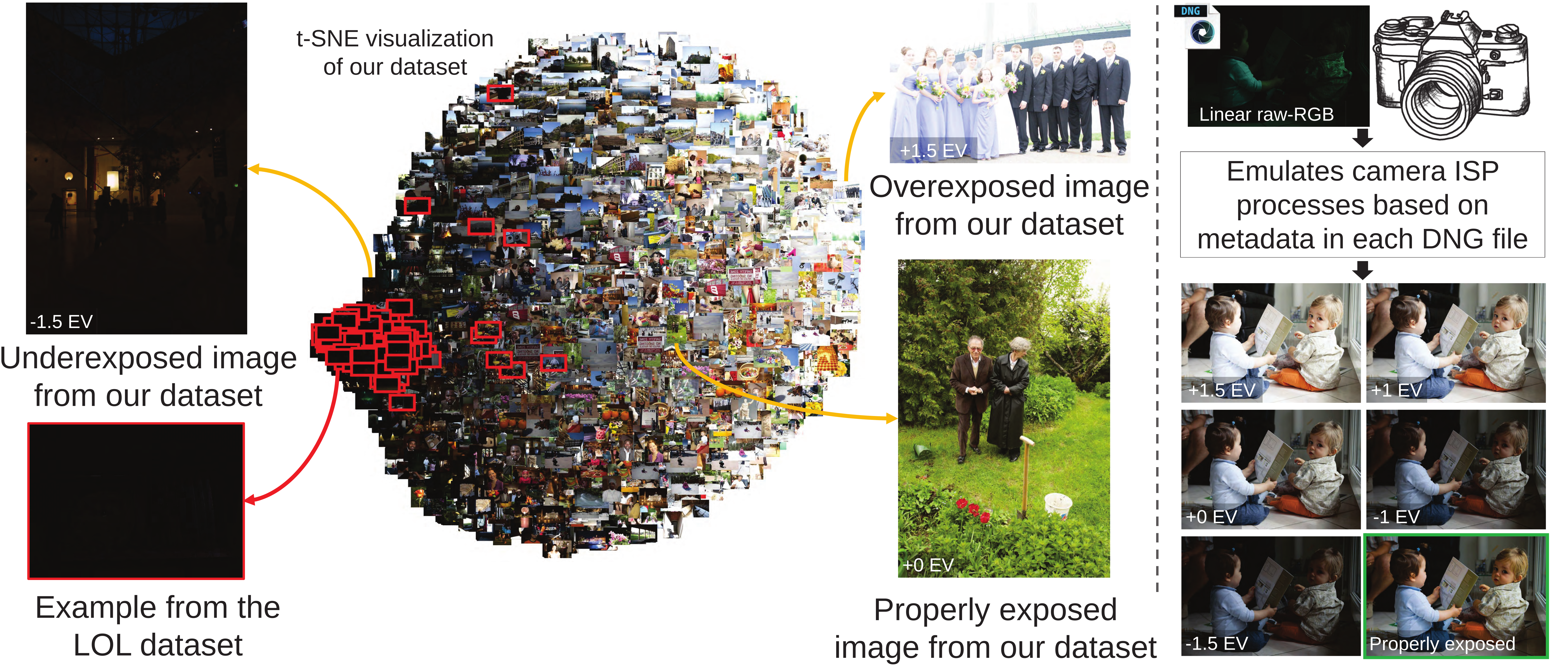}
\vspace{-6mm}
\caption{Dataset overview. Our dataset contains images with different exposure error types and their corresponding properly exposed reference images. Shown is a t-SNE visualization \cite{maaten2008visualizing} of all images in our dataset and the low-light (LOL) paired dataset (outlined in red) \cite{Chen2018Retinex}.  Notice that LOL covers a relatively small fraction of the possible exposure levels, as compared
to our introduced dataset.
Our dataset was rendered from linear raw-RGB images taken from the MIT-Adobe FiveK dataset \cite{fivek}. Each image was rendered with different relative exposure values (EVs) by an accurate emulation of the camera ISP processes.\vspace{-4mm}}
\label{fig:dataset}
\end{figure}

\noindent\textbf{Paired Dataset } Paired datasets are crucial for supervised learning for image enhancement tasks. Existing paired datasets for exposure correction focus only on low-light underexposed images. Representative examples include Wang et al.'s dataset \cite{DeepUPE} and the low-light (LOL) paired dataset \cite{Chen2018Retinex}.
Unlike existing datasets for exposure correction, we introduce a large image dataset rendered with a wide range of exposure errors. Fig.\ \ref{fig:dataset} shows a comparison between our dataset and the LOL dataset in terms of the number of images and the variety of exposure errors in each dataset. The LOL dataset covers a relatively small fraction of the possible exposure levels, as compared to our introduced dataset.  Our dataset is based on the MIT-Adobe FiveK dataset~\cite{fivek} and is accurately rendered by adjusting the high tonal values provided in camera sensor raw-RGB images to realistically emulate camera exposure errors.  An alternative worth noting is to use a large HDR dataset to produce training data---for example, the Google HDR+ dataset~\cite{hasinoff2016burst}. One drawback, however, is that this dataset is a composite of a varying number of smartphone captured raw-RGB images that were first aligned to a composite raw-RGB image.  The target ground truth image is based on an HDR-to-LDR algorithm applied to this composite raw-RGB image~\cite{hasinoff2016burst, HDRNET}.  We opt instead to use the FiveK dataset as it starts with a single high-quality raw-RGB image and the ground truth result is generated by an expert photographer.

\begin{figure}[t]
\centering
\includegraphics[width=1\linewidth]{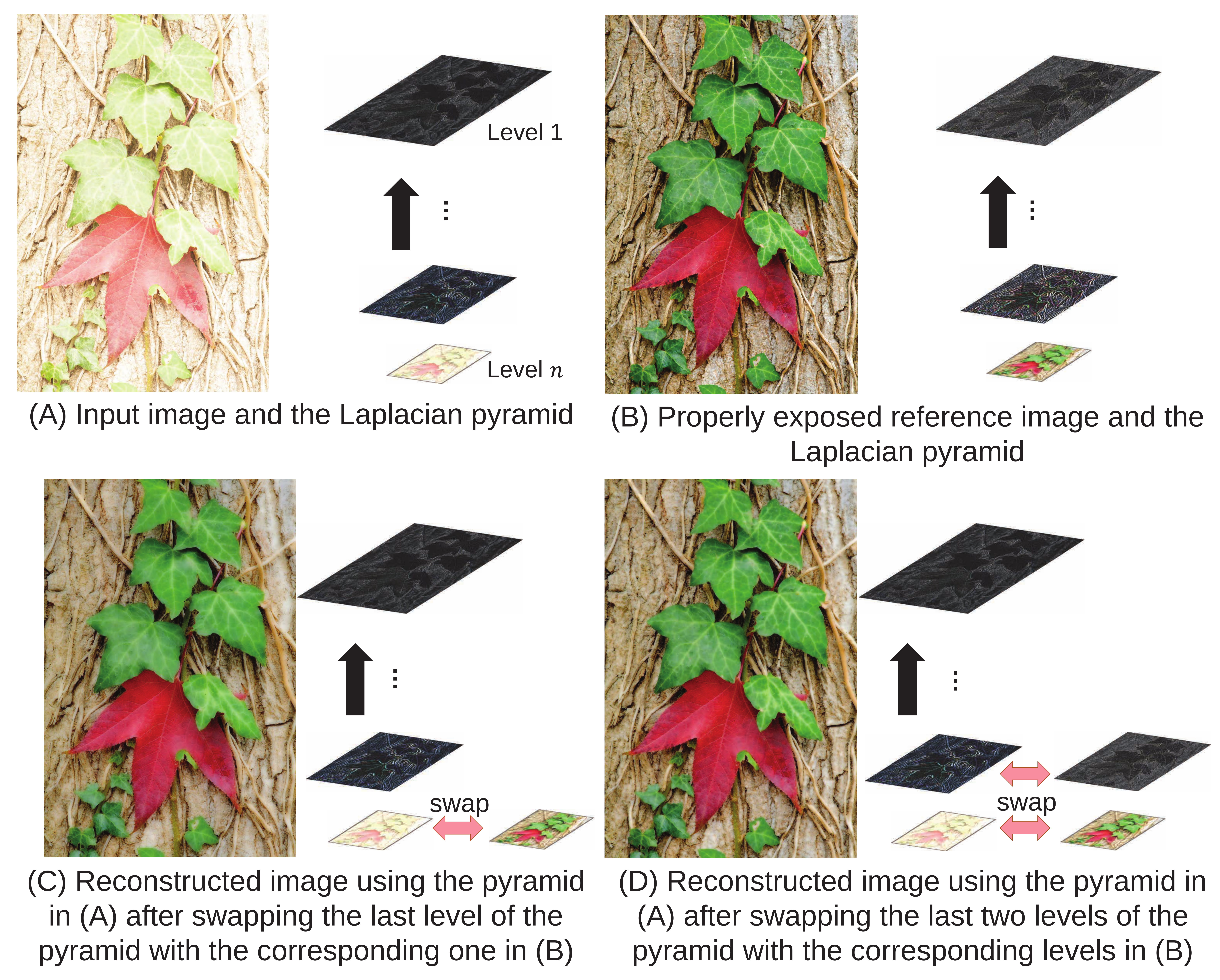}
\vspace{-6mm}
\caption{Motivation behind our coarse-to-fine exposure correction approach. Example of an overexposed image and its corresponding properly exposed image shown in (A) and (B), respectively. The Laplacian pyramid decomposition allows us to enhance the color and detail information sequentially, as shown in (C) and (D), respectively.\vspace{-4mm}}
\label{fig:idea}
\end{figure}

\section{Our Dataset} \label{subsec:data}

To train our model, we need a large number of training images rendered with realistic over-  and underexposure errors and corresponding properly exposed ground truth images.
As discussed in Sec.~\ref{sec:relatedwork}, such datasets are currently not publicly available to support exposure correction research. For this reason, our first task is to create a new dataset.  Our dataset is rendered from the MIT-Adobe FiveK dataset~\cite{fivek}, which has 5,000 raw-RGB images and corresponding sRGB images rendered manually by five expert photographers~\cite{fivek}.

For each raw-RGB image, we use the Adobe Camera Raw SDK~\cite{CameraRaw} to emulate different EVs as would be applied by a camera~\cite{schewe2010real}. Adobe Camera Raw accurately emulates the nonlinear camera rendering procedures using metadata embedded in each DNG raw file \cite{afifi2019color, schewe2010real}. We render each raw-RGB image with different digital EVs to mimic real exposure errors. Specifically, we use the  relative EVs $-1.5$, $-1$, $+0$, $+1$, and $+1.5$ to render images with underexposure errors, a zero gain of the original EV, and overexposure errors, respectively. The zero-gain relative EV is equivalent to the original exposure settings applied onboard the camera during capture time.

As the ground truth images, we use images that were manually retouched by an expert photographer (referred to as Expert C in \cite{fivek}) as our target correctly exposed images, rather than using our rendered images with $+0$ relative EV. The reason behind this choice is that a significant number of images contain backlighting or partial exposure errors in the original exposure capture settings. The expert adjusted images were performed in ProPhoto RGB color space \cite{fivek} (rather than raw-RGB), which we converted to a standard 8-bit sRGB color space encoding.

In total, our dataset contains 24,330 8-bit sRGB images with different digital exposure settings. We discarded a small number of images that had misalignment with their corresponding ground truth image. These misalignments are due to different usage of the DNG crop area metadata by Adobe Camera Raw SDK and the expert. Our dataset is divided into three sets: (i) training set of 17,675 images, (ii) validation set of 750 images, and (iii) testing set of 5,905 images.   The training, validation, and testing sets, use different images taken from the FiveK dataset. This means the training, validation, and testing images do not share any images in common. Fig.\ \ref{fig:dataset} shows examples of our generated 8-bit sRGB images and the corresponding properly exposed 8-bit sRGB reference images.

\begin{figure*}[t]
\centering
\includegraphics[width=\linewidth]{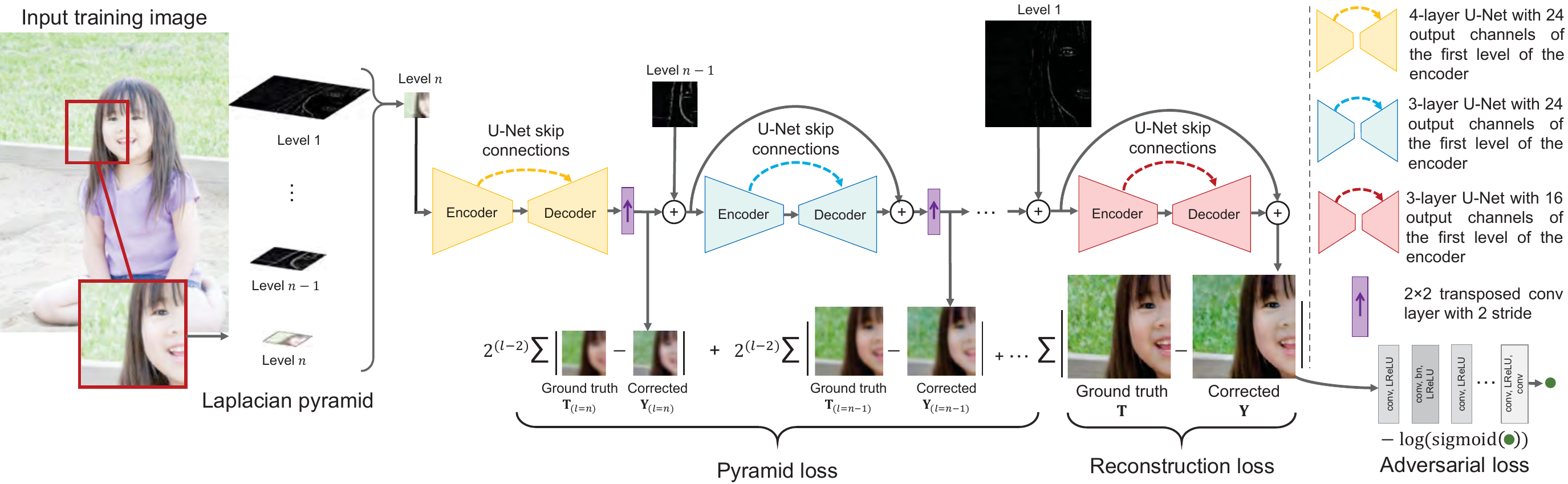}
\vspace{-6mm}
\caption{Overview of our image exposure correction architecture. We propose a coarse-to-fine deep network to progressively correct exposure errors in 8-bit sRGB images.
Our network first corrects the global color captured at the final level of the Laplacian pyramid and then the subsequent frequency layers.\vspace{-4mm}}
\label{fig:main}
\end{figure*}

\section{Our Method} \label{sec:method}
Given an 8-bit sRGB input image, $\mathbf{I}$, rendered with the incorrect exposure setting, our method aims to produce an output image, $\mathbf{Y}$, with fewer exposure errors than those in $\mathbf{I}$. As we simultaneously target both over- and underexposed errors, our input image, $\mathbf{I}$, is expected to contain regions of nearly over- or under-saturated values with corrupted color and detail information.
We propose to correct color and detail errors of $\mathbf{I}$ in a sequential manner. Specifically, we process a multi-resolution representation of $\mathbf{I}$, rather than directly dealing with the original form of $\mathbf{I}$. We use the Laplacian pyramid \cite{burt1983laplacian} as our multiresolution decomposition, which is derived  from the Gaussian pyramid of $\mathbf{I}$.

\subsection{Coarse-to-Fine Exposure Correction}

Let $\mathbf{X}$ represent the Laplacian pyramid of $\mathbf{I}$ with $n$ levels, such that  $\mathbf{X}_{(l)}$ is the $l^\texttt{th}$ level of $\mathbf{X}$. The last level of this pyramid (i.e., $\mathbf{X}_{(n)}$) captures low-frequency information of $\mathbf{I}$, while the first level (i.e., $\mathbf{X}_{(1)}$) captures the high-frequency information. Such frequency levels can be categorized into: (i) global color information of $\mathbf{I}$ stored in the low-frequency level and (ii) image coarse-to-fine details stored in the mid- and high-frequency levels. These levels can be later used to reconstruct the full-color image $\mathbf{I}$.

Fig.\ \ref{fig:idea} motivates our coarse-to-fine approach to exposure correction.  Figs.\ \ref{fig:idea}-(A) and (B) show an example overexposed image and its corresponding well-exposed target, respectively.
As observed, a significant exposure correction can be obtained by  using only the low-frequency layer (i.e., the global color information) of the target image in the Laplacian pyramid reconstruction process, as shown in Fig.~\ref{fig:idea}-(C). We can then improve the final image by enhancing the details in a sequential way by correcting each level of the Laplacian pyramid, as shown in Fig.~\ref{fig:idea}-(D). Practically, we do not have access to the properly exposed image in Fig.~\ref{fig:idea}-(B) at the inference stage, and thus our goal is to predict the missing color/detail information of each level in the Laplacian pyramid.

Inspired by this observation and the success of coarse-to-fine architectures for various other computer vision tasks (e.g., \cite{denton2015deep, shaham2019singan, lai2017deep, ma2020efficient}), we design a DNN that corrects the global color and detail information of $\mathbf{I}$ in a sequential manner using the Laplacian pyramid decomposition.
The remaining parts of this section explain the technical details of our model (Sec.\ \ref{subsec:network}), including details of the losses (Sec.\ \ref{subsec:losses}),
inference phase (Sec.\ \ref{subsec:Inference}), and training (Sec.\ \ref{subsec:training_details}).

\subsection{Coarse-to-Fine Network} \label{subsec:network}
Our image exposure correction architecture sequentially processes the $n$-level Laplacian pyramid, $\mathbf{X}$, of the input image, $\mathbf{I}$, to produce the final corrected image, $\mathbf{Y}$. The proposed model consists of $n$ sub-networks. Each of these sub-networks is a U-Net-like architecture \cite{unet} with untied weights. We allocate the network capacity in the form of weights based on how significantly each sub-problem (i.e., global color correction and detail enhancement) contributes to our final result.
Fig.\ \ref{fig:main} provides an overview of our network. As shown, the largest (in terms of weights) sub-network in our architecture is dedicated to processing the global color information in $\mathbf{I}$ (i.e., $\mathbf{X}_{(n)}$). This sub-network (shown in yellow in Fig.\ \ref{fig:main}) processes the low-frequency level $\mathbf{X}_{(n)}$ and produces an upscaled image $\mathbf{Y}_{(n)}$. The upscaling process scales up the output of our sub-network by a factor of two using strided transposed convolution with trainable weights. Next, we add the first mid-frequency level $\mathbf{X}_{(n-1)}$ to $\mathbf{Y}_{(n)}$ to be processed by the second sub-network in our model. This sub-network enhances the corresponding details of the current level and produces a residual layer that is then added to $\mathbf{Y}_{(n)} + \mathbf{X}_{(n-1)}$ to reconstruct image $\mathbf{Y}_{(n-1)}$, which is equivalent to the corresponding Gaussian pyramid level $n-1$. This refinement-upsampling process proceeds until the final output image, $\mathbf{Y}$, is produced. Our network is fully differentiable and thus can be trained in an end-to-end manner.
Additional details of our network are provided in the supplementary materials.
The code and weights for our model will be released to support reproducibility and facilitate future research.

\subsection{Losses}\label{subsec:losses}

We train our model end-to-end to minimize the following loss function:
\begin{equation}
\label{eq:final_loss}
\mathcal{L} = \mathcal{L}_{\text{rec}} + \mathcal{L}_{\text{pyr}} + \mathcal{L}_{\text{adv}},
\end{equation}
where  $\mathcal{L}_{\text{rec}}$ denotes the reconstruction loss, $\mathcal{L}_{\text{pyr}}$ the pyramid loss, and $\mathcal{L}_{\text{adv}}$ the adversarial loss. The individual losses are defined next.

\paragraph{Reconstruction Loss:}
We use the $\texttt{L}_1$ loss function between the reconstructed and properly exposed reference images. This loss can be expressed as follows:
\begin{equation}
\label{eq:wo_pyramid_loss}
\mathcal{L}_{\text{rec}} = \sum_{p=1}^{3hw} \left|\mathbf{Y}(p) - \mathbf{T}(p)\right|,
\end{equation}
where $h$ and $w$ denote the height and width of the training image, respectively, and
$p$ is the index of each pixel in our corrected image, $\mathbf{Y}$, and the
corresponding properly exposed reference image, $\mathbf{T}$, respectively.

\begin{figure}[b]
\centering
\includegraphics[width=\linewidth]{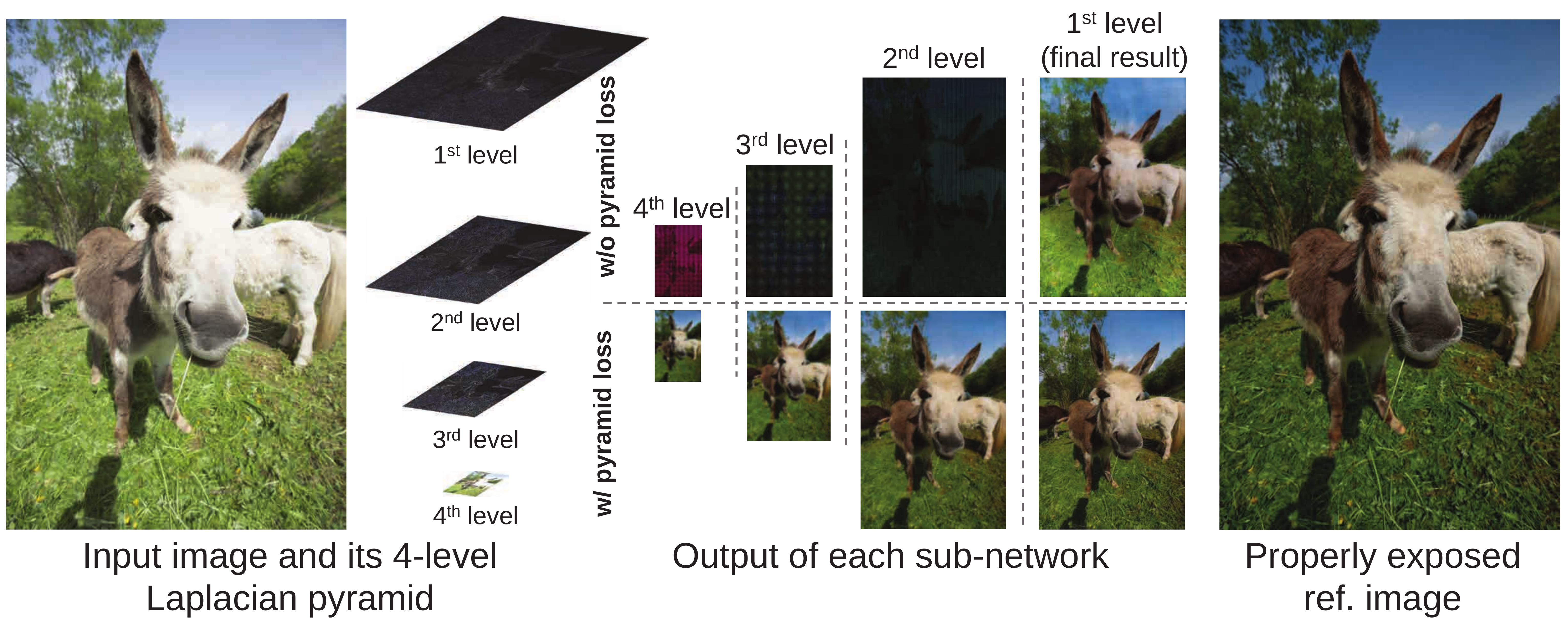}
\vspace{-6mm}
\caption{Multiscale losses. Shown are the output of each sub-net trained with and without the pyramid loss (Eq.\ \ref{eq:w_pyramid_loss}).}
\label{fig:how_it_works}
\end{figure}

\paragraph{Pyramid Loss:}
To guide each sub-network to follow the Laplacian pyramid reconstruction procedure, we introduce dedicated losses at each pyramid level.  Let $\mathbf{T}_{(l)}$ denote the  $l^{\texttt{th}}$ level of the Gaussian pyramid of our reference image, $\mathbf{T}$, after upsampling by a factor of two. We use a simple interpolation process for the upsampling operation \cite{mertens2009exposure}. Our pyramid loss is computed as follows:
\begin{equation}
\label{eq:w_pyramid_loss}
\mathcal{L}_{\text{pyr}} = \sum_{l=2}^{n}2^{(l-2)}\sum_{p=1}^{3h_lw_l} \left|\mathbf{Y}_{(l)}(p) - \mathbf{T}_{(l)}(p)\right|,
\end{equation}
where $h_l$ and $w_l$ are twice the height and width of the $l^{\texttt{th}}$ level in the Laplacian pyramid of the training image, respectively, and
$p$ is the index of each pixel in our corrected image at the $l^{\texttt{th}}$ level $\mathbf{Y}_{(l)}$ and the
properly exposed reference image at the same level $\mathbf{T}_{(l)}$, respectively. The pyramid loss not only gives a principled interpretation of the task of each sub-network but also results in less visual artifacts compared to training using only the reconstruction loss, as can be seen in Fig.~\ref{fig:how_it_works}. Notice that without the intermediate pyramid losses, the multi-scale reconstructions, shown in Fig.\ \ref{fig:how_it_works} (right-top), deviate widely from the intermediate Gaussian targets compared to using the pyramid loss at each scale, as shown in Fig.\ \ref{fig:how_it_works} (right-bottom). We provide supporting justification for this loss with an ablation study in the supplementary materials.

\paragraph{Adversarial Loss:} To perceptually enhance the reconstruction of the corrected image output in terms of realism and appeal, we also consider an adversarial loss as a regularizer.
This adversarial loss term can be described by the following equation \cite{goodfellow2014generative}:

\begin{equation}
\label{eq:w_adv_loss}
\mathcal{L}_{\text{adv}} = -3hwn \log\left(\mathcal{S}\left(\mathcal{D}\left(\mathbf{Y}\right)\right)\right),
\end{equation}
where $\mathcal{S}$ is the sigmoid function and $\mathcal{D}$ is a discriminator DNN that is trained together with our main network. We provide the details of our discriminator network and visual comparisons between our results using non-adversarial and adversarial training in the supplementary materials.

\subsection{Inference Stage}\label{subsec:Inference}

Our network is fully convolutional and can process input images with different resolutions. While our model requires a reasonable memory size ($\sim$7M parameters), processing high-resolution images requires a high computational power that may not always be available. Furthermore, processing images with considerably higher resolution (e.g., 16-megapixel) than the range of resolutions used in the training process can affect our model's robustness with large homogeneous image regions. This issue arises because our network was trained on a certain range of effective receptive fields, which is very low compared to the receptive fields required for images with very high resolution. To that end, we use the bilateral guided upsampling method \cite{chen2016bilateral} to process high-resolution images. First, we resize the input test image to have a maximum dimension of 512 pixels. Then, we process the downsampled version of the input image using our model, followed by applying the fast upsampling technique~\cite{chen2016bilateral} with a bilateral grid of $22\!\times\!22\!\times\!8$ cells. This process allows us to process a 16-megapixel image in $\sim$4.5 seconds on average. This time includes $\sim$0.5 seconds to run our network on an NVIDIA$^\circledR$ GeForce GTX 1080$^\texttt{TM}$ GPU and $\sim$4 seconds on an Intel$^\circledR$ Xeon$^\circledR$ E5-1607 @ 3.10 GHz machine for the guided upsampling process. Note the runtime of the guided upsampling step can be significantly improved with a Halide implementation \cite{HALIDE}.

\begin{figure}[t]
\centering
\includegraphics[width=\linewidth]{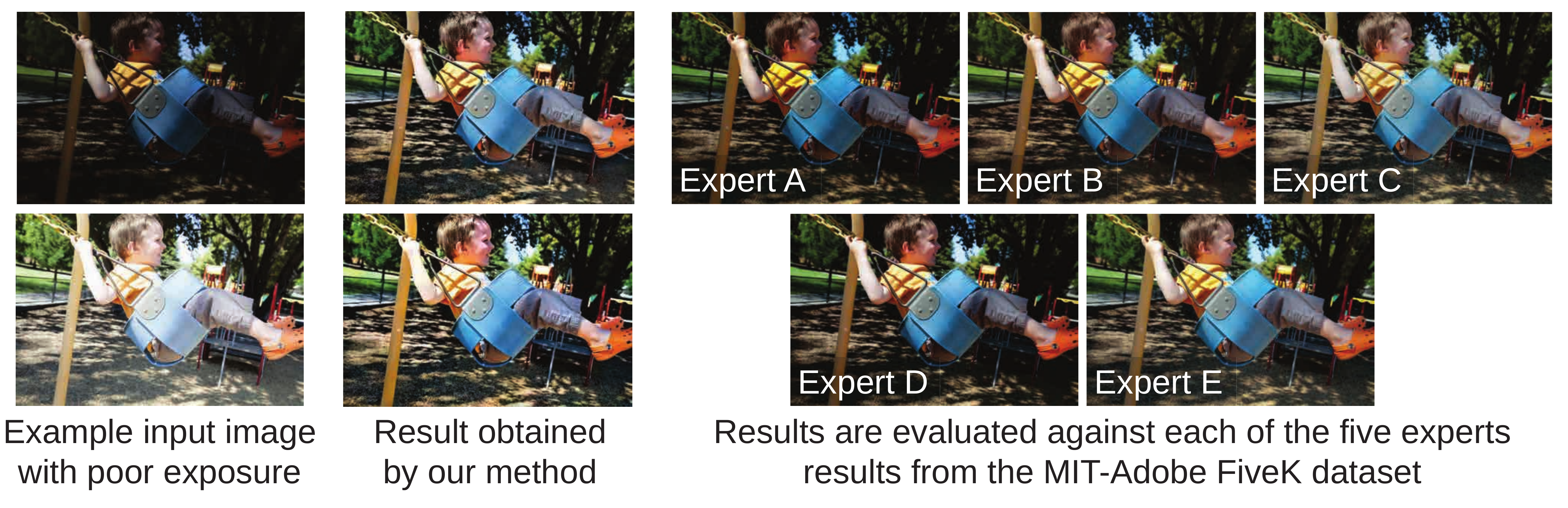}
\vspace{-6mm}
\caption{We evaluate the results of input images against all five expert photographers' edits from the FiveK dataset~\cite{fivek}.\vspace{-4mm}}
\label{fig:experts}
\end{figure}

\begin{figure*}
\centering
\includegraphics[width=\linewidth]{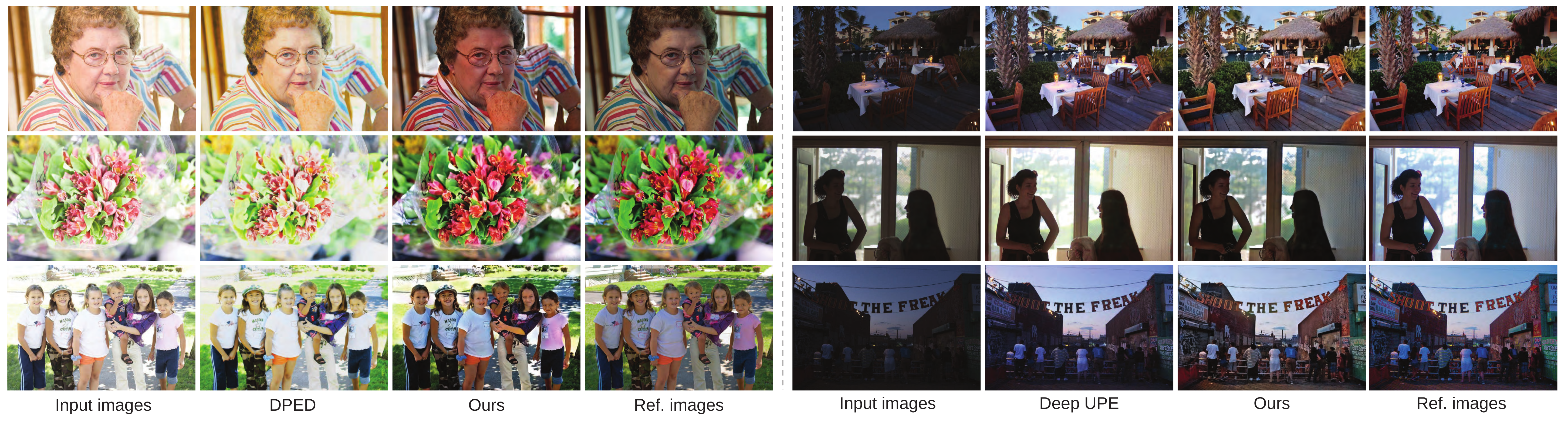}
\vspace{-7mm}
\caption{Qualitative results of correcting images with exposure errors. Shown are the input images from our test set, results from the DPED \cite{DPED}, results from the Deep UPE \cite{DPE}, our results, and the corresponding ground truth images.\vspace{-2mm}}
\label{fig:qualitative_our_set}
\end{figure*}

\subsection{Training Details}\label{subsec:training_details}

In our implementation, we use a Laplacian pyramid with four levels (i.e., $n=4$) and thus we have four sub-networks in our model---an ablation study evaluating the effect on the
number of Laplacian levels, including a comparison with a vanilla U-Net architecture, is presented in the supplementary materials. We trained our model on patches randomly extracted from training images with different dimensions. We first train on patches of size $128\!\times\!128$ pixels. Next, we continue training on $256\!\times\!256$ patches, followed by training on $512\!\times\!512$ patches. We use the Adam optimizer \cite{kingma2014adam} to minimize our loss function in Eq.\ \ref{eq:final_loss}.
Inspired by previous work \cite{ma2017pose}, we initially train
 without the adversarial loss term $\mathcal{L}_{\text{adv}}$ to speed up the convergence of our main network. Upon convergence, we then add the adversarial loss term $\mathcal{L}_{\text{adv}}$
 and fine-tune our network to enhance our initial results. Additional training details are provided  in the supplementary materials.

\section{Empirical Evaluation} \label{sec:results}

We compare our method against a broad range of existing methods for exposure correction and image enhancement. We first present quantitative results and comparisons in Sec.\ \ref{subsec:quan-results}, followed by qualitative comparisons in Sec.\ \ref{subsec:qual-results}.

\subsection{Quantitative Results} \label{subsec:quan-results}

To evaluate our method, we use our test set, which consists of 5,905 images rendered with different exposure settings, as described in Sec.\ \ref{subsec:data}. Specifically, our test set includes 3,543 well-exposed/overexposed images rendered with $+0$, $+1$, and $+1.5$ relative EVs, and 2,362 underexposed images rendered with $-1$ and $-1.5$ relative EVs.

We adopt the following three standard metrics to evaluate the pixel-wise accuracy and the perceptual quality of our results: (i)  peak signal-to-noise ratio (PSNR), (ii)  structural similarity index measure (SSIM) \cite{wang2004image}, and (iii)  perceptual index ($\texttt{PI}$) \cite{blau20182018}. The $\texttt{PI}$ is given by:
\begin{equation}
\label{eq:PI}
\texttt{PI} = 0.5 (10 - \texttt{Ma} + \texttt{NIQE}),
\end{equation}
where both $\texttt{Ma}$ \cite{ma2017learning} and $\texttt{NIQE}$  \cite{mittal2012making} are \textit{no-reference} image quality metrics.

\begin{table*}
\caption{Quantitative evaluation on our introduced test set. \textbf{The best results are highlighted with green and bold. The second- and third-best results are highlighted in yellow and red, respectively.} We compare each method with properly exposed reference image sets rendered by five expert photographers \cite{fivek}.  For each method,
we present peak signal-to-noise ratio (PSNR), structural similarity index measure (SSIM) \cite{wang2004image}, and perceptual index (PI) \cite{blau20182018}.  We denote methods designed for underexposure correction in gray. Non-deep learning methods are marked by $\textasteriskcentered$. The terms U and S stand for unsupervised and supervised, respectively. Notice that higher PSNR and SSIM values are better, while lower PI values indicate better perceptual quality.
\vspace{-6mm}}\label{table:results}
\begin{center}
\scalebox{0.67}{
\begin{tabular}{|l|c|c|c|c|c|c|c|c|c|c|c|c|c|}
\cline{1-14}
\multirow{2}{*}{Method} & \multicolumn{2}{c|}{Expert A} & \multicolumn{2}{c|}{Expert B} & \multicolumn{2}{c|}{Expert C} & \multicolumn{2}{c|}{Expert D} & \multicolumn{2}{c|}{Expert E} & \multicolumn{2}{c|}{Avg.} & \multirow{2}{*}{$\texttt{PI}$} \\ \cline{2-13}
 & PSNR & SSIM & PSNR & SSIM & PSNR & SSIM & PSNR & SSIM & PSNR & SSIM & PSNR & SSIM & \\ \cline{1-14}

\multicolumn{14}{|c|}{\cellcolor[HTML]{CCECEB}$+0$, $+1$, and $+1.5$ relative EVs (3,543 properly exposed and overexposed images)}\\ \hline
HE \cite{10.5555/559707} $\textasteriskcentered$& 16.140 & \cellcolor[HTML]{FFCCCB}0.686 & 16.277 & 0.672 & 16.531 & 0.699 & 16.643 & 0.669 & 17.321  & 0.691 & 16.582 & 0.683 & 2.351\\
CLAHE \cite{adaptivehisteq} $\textasteriskcentered$& 13.934 & 0.568 & 14.689 & 0.586 & 14.453 &  0.584 & 15.116 & 0.593 & 15.850 & 0.612 & 14.808 & 0.589 & 2.270\\
WVM \cite{fu2016weighted} $\textasteriskcentered$& 12.355 & 0.624 & 13.147 & 0.656 & 12.748 & 0.645 & 14.059 & 0.669 & 15.207 & 0.690 & 13.503 & 0.657 & 2.342\\
\cellcolor[HTML]{D5D5D5}LIME \cite{guo2016lime, guo2017lime} $\textasteriskcentered$& 09.627 & 0.549 & 10.096 & 0.569 & 9.875 & 0.570 & 10.936 & 0.597 & 11.903 & 0.626 & 10.487 & 0.582 & 2.412\\
HDR CNN \cite{HDRCNN} w/ RHT \cite{yang2018image}& 13.151 & 0.475 & 13.637 & 0.478 & 13.622 & 0.497 & 14.177 & 0.479 & 14.625 & 0.503 & 13.842 & 0.486 & 4.284\\
HDR CNN \cite{HDRCNN} w/ PS \cite{dayley2010photoshop}& 14.804  & 0.651 & 15.622 & 0.689 & 15.348 & 0.670 & 16.583  & 0.685 & 18.022 & \cellcolor[HTML]{FFCCCB}0.703 & 16.076 & 0.680 & \cellcolor[HTML]{FFCCCB}2.248\\
DPED (iPhone) \cite{DPED}& 12.680  &  0.562 & 13.422 & 0.586 & 13.135 & 0.581 & 14.477 & 0.596 & 15.702 & 0.630 & 13.883 & 0.591 & 2.909 \\
DPED (BlackBerry) \cite{DPED} & 15.170 & 0.621 & 16.193 & 0.691 &  15.781 & 0.642 & 17.042 & 0.677 & 18.035 & 0.678 & 16.444 & 0.662 & 2.518\\
DPED (Sony) \cite{DPED}& \cellcolor[HTML]{FFCCCB}16.398 & 0.672 & \cellcolor[HTML]{FFCCCB}17.679 & \cellcolor[HTML]{FFCCCB}0.707 & \cellcolor[HTML]{FFCCCB}17.378 & 0.697 & \cellcolor[HTML]{FFCCCB}17.997 & \cellcolor[HTML]{FFCCCB}0.685 & \cellcolor[HTML]{FFCCCB}18.685 & 0.700 & \cellcolor[HTML]{FFCCCB}17.627 & \cellcolor[HTML]{FFCCCB}0.692 & 2.740\\
DPE (HDR) \cite{DPE} & 14.399 & 0.572 & 15.219 & 0.573 & 15.091 & 0.593 & 15.692 & 0.581 & 16.640 & 0.626 & 15.408 & 0.589 & 2.417\\
DPE (U-FiveK) \cite{DPE} & 14.314 & 0.615 & 14.958 & 0.628 & 15.075 & 0.645 & 15.987 & 0.647 & 16.931 & 0.667 & 15.453 & 0.640 & 2.630\\
DPE (S-FiveK) \cite{DPE} & 14.786 & 0.638 & 15.519 & 0.649 & 15.625 &  0.668 & 16.586 & 0.664 &  17.661 & 0.684 & 16.035 & 0.661 & 2.621\\
\cellcolor[HTML]{D5D5D5}HQEC \cite{HQEC} $\textasteriskcentered$& 11.775 & 0.607 & 12.536 & 0.631 & 12.127 & 0.627 & 13.424 & 0.652 & 14.511 & 0.675 & 12.875 & 0.638 & 2.387\\
\cellcolor[HTML]{D5D5D5}RetinexNet \cite{Chen2018Retinex} & 10.149 & 0.570 & 10.880 & 0.586 & 10.471 & 0.595 & 11.498 & 0.613 & 12.295 & 0.635 & 11.059 & 0.600 & 2.933\\
\cellcolor[HTML]{D5D5D5}Deep UPE \cite{DeepUPE} & 10.047 & 0.532 & 10.462 & 0.568 & 10.307 & 0.557 & 11.583 & 0.591 & 12.639 & 0.619 & 11.008 & 0.573 & 2.428\\
\cellcolor[HTML]{D5D5D5}Zero-DCE \cite{guo2020zero}  & 10.116 & 0.503 & 10.767 & 0.502 & 10.395 & 0.514 & 11.471 & 0.522 & 12.354 & 0.557 & 11.0206 & 0.5196 & 2.774 \\
\hdashline
Our method w/o $\mathcal{L}_{\text{adv}}$ & \cellcolor[HTML]{79CC7A}\textbf{18.976} & \cellcolor[HTML]{79CC7A}\textbf{0.743} & \cellcolor[HTML]{79CC7A}\textbf{19.767} & \cellcolor[HTML]{79CC7A}\textbf{0.731} & \cellcolor[HTML]{79CC7A}\textbf{19.980} & \cellcolor[HTML]{79CC7A}\textbf{0.768} & \cellcolor[HTML]{79CC7A}\textbf{18.966} & \cellcolor[HTML]{79CC7A}\textbf{0.716} & \cellcolor[HTML]{79CC7A}\textbf{19.056} & \cellcolor[HTML]{79CC7A}\textbf{0.727} & \cellcolor[HTML]{79CC7A}\textbf{19.349} & \cellcolor[HTML]{79CC7A}\textbf{0.737} & \cellcolor[HTML]{FFFBA3}2.189\\
Our method w/ $\mathcal{L}_{\text{adv}}$ & \cellcolor[HTML]{FFFBA3}18.874 & \cellcolor[HTML]{FFFBA3}0.738 & \cellcolor[HTML]{FFFBA3}19.569 & \cellcolor[HTML]{FFFBA3}0.718 & \cellcolor[HTML]{FFFBA3}19.788 & \cellcolor[HTML]{FFFBA3}0.760 & \cellcolor[HTML]{FFFBA3}18.823 & \cellcolor[HTML]{FFFBA3}0.705 & \cellcolor[HTML]{FFFBA3}18.936 & \cellcolor[HTML]{FFFBA3}0.719 & \cellcolor[HTML]{FFFBA3}19.198 & \cellcolor[HTML]{FFFBA3}0.728 & \cellcolor[HTML]{79CC7A}\textbf{2.183}\\\hline

\multicolumn{14}{|c|}{\cellcolor[HTML]{CCECEB}$-1$ and $-1.5$ relative EVs (2,362 underexposed images)}\\ \hline
HE \cite{10.5555/559707} $\textasteriskcentered$& 16.158 & 0.683 & 16.293 & 0.669 & 16.517  & 0.692 & 16.632 &  0.665 & 17.280 & 0.684 & 16.576 & 0.679 & 2.486\\
CLAHE \cite{adaptivehisteq} $\textasteriskcentered$& 16.310 & 0.619 & 17.140 & 0.646 & 16.779 & 0.621 & 15.955 & 0.613 & 15.568  & 0.608 & 16.350 & 0.621 & 2.387\\
WVM \cite{fu2016weighted} $\textasteriskcentered$& 17.686 & 0.728 & 19.787 & \cellcolor[HTML]{79CC7A}\textbf{0.764} & 18.670 & 0.728 & 18.568 & \cellcolor[HTML]{FFFBA3}0.729 & 18.362 & \cellcolor[HTML]{FFCCCB}0.724 & 18.615 & 0.735 & 2.525\\
\cellcolor[HTML]{D5D5D5}LIME \cite{guo2016lime, guo2017lime} $\textasteriskcentered$& 13.444 & 0.653 & 14.426 & 0.672 & 13.980 & 0.663 & 15.190 & 0.673 & 16.177 & 0.694 & 14.643 & 0.671 & 2.462\\
HDR CNN \cite{HDRCNN} w/ RHT \cite{yang2018image}& 14.547 & 0.456 & 14.347 & 0.427 & 14.068 & 0.441 & 13.025 & 0.398 &  11.957 & 0.379 & 13.589 & 0.420 & 5.072\\
HDR CNN \cite{HDRCNN} w/ PS \cite{dayley2010photoshop}& 17.324 & 0.692 & 18.992 & 0.714 & 18.047 & 0.696 & 18.377 & 0.689 & \cellcolor[HTML]{FFFBA3}19.593 & 0.701 & 18.467 & 0.698 & \cellcolor[HTML]{79CC7A}\textbf{2.294}\\
DPED (iPhone) \cite{DPED}& 18.814 & 0.680 & \cellcolor[HTML]{FFFBA3}21.129 & 0.712 & 20.064 & 0.683 &  \cellcolor[HTML]{79CC7A}\textbf{19.711}  & 0.675 & \cellcolor[HTML]{FFCCCB}19.574  & 0.676 & \cellcolor[HTML]{FFFBA3}19.858 & 0.685 & 2.894\\
DPED (BlackBerry) \cite{DPED} & \cellcolor[HTML]{FFFBA3}19.519 & 0.673 & \cellcolor[HTML]{79CC7A}\textbf{22.333} & 0.745 & 20.342 & 0.669 & 19.611 & 0.683 & 18.489 & 0.653 & \cellcolor[HTML]{79CC7A}\textbf{20.059} & 0.685 & 2.633\\
DPED (Sony) \cite{DPED}& 18.952 & 0.679 & 20.072 & 0.691 & 18.982 &  0.662 & 17.450 & 0.629 & 15.857 & 0.601 & 18.263 & 0.652 & 2.905\\
DPE (HDR) \cite{DPE} & 17.625 & 0.675 & 18.542 & 0.705 & 18.127  & 0.677 & 16.831 & 0.665 & 15.891 & 0.643 & 17.403 & 0.673 & \cellcolor[HTML]{FFFBA3}2.340\\
DPE (U-FiveK) \cite{DPE} & 19.130 & 0.709 & 19.574 & 0.674 & 19.479 & 0.711 & 17.924 & 0.665 & 16.370 & 0.625 & 18.495 & 0.677 & 2.571\\
DPE (S-FiveK) \cite{DPE} & \cellcolor[HTML]{79CC7A}\textbf{20.153} & \cellcolor[HTML]{FFCCCB}0.738 & \cellcolor[HTML]{FFCCCB}20.973 & 0.697 & \cellcolor[HTML]{79CC7A}\textbf{20.915} & 0.738 & \cellcolor[HTML]{FFCCCB}19.050 & 0.688 & 17.510 & 0.648 & \cellcolor[HTML]{FFCCCB}19.720 & 0.702  & 2.564\\
\cellcolor[HTML]{D5D5D5}HQEC \cite{HQEC} $\textasteriskcentered$& 15.801 & 0.692 & 17.371 & 0.718 & 16.587 & 0.700 & 17.090 & 0.705 & 17.675 & 0.716 & 16.905 & 0.706 & 2.532\\
\cellcolor[HTML]{D5D5D5}RetinexNet \cite{Chen2018Retinex}  & 11.676 & 0.607 & 12.711 & 0.611 & 12.132 & 0.621 & 12.720 & 0.618 & 13.233 & 0.637 & 12.494 & 0.619 & 3.362\\
\cellcolor[HTML]{D5D5D5}Deep UPE \cite{DeepUPE} & 17.832 & 0.728 & 19.059 & \cellcolor[HTML]{FFFBA3}0.754 & 18.763 & \cellcolor[HTML]{FFCCCB}0.745 & \cellcolor[HTML]{FFFBA3}19.641 & \cellcolor[HTML]{79CC7A}\textbf{0.737} & \cellcolor[HTML]{79CC7A}\textbf{20.237} & \cellcolor[HTML]{79CC7A}\textbf{0.740} & 19.106 & \cellcolor[HTML]{FFFBA3}0.741 & 2.371\\ 
\cellcolor[HTML]{D5D5D5}Zero-DCE \cite{guo2020zero}  & 13.935 & 0.585 & 15.239 & 0.593 & 14.552 & 0.589 & 15.202 & 0.587 & 15.893 & 0.614 & 14.9642 & 0.5936 & 3.001 \\
\hdashline
Our method w/o $\mathcal{L}_{\text{adv}}$ & \cellcolor[HTML]{FFCCCB}19.432 & \cellcolor[HTML]{FFFBA3}0.750 & 20.590 & \cellcolor[HTML]{FFCCCB}0.739 & \cellcolor[HTML]{FFFBA3}20.542 & \cellcolor[HTML]{79CC7A}\textbf{0.770} & 18.989 & \cellcolor[HTML]{FFCCCB}0.723 & 18.874 & \cellcolor[HTML]{FFFBA3}0.727 & 19.685 & \cellcolor[HTML]{79CC7A}\textbf{0.742} & 2.344 \\
Our method w/ $\mathcal{L}_{\text{adv}}$ & 19.475 & \cellcolor[HTML]{79CC7A}\textbf{0.751} & 20.546 & 0.730 & \cellcolor[HTML]{FFCCCB}20.518 & \cellcolor[HTML]{FFFBA3}0.768 & 18.935 & 0.715 & 18.756 & 0.719 & 19.646 & \cellcolor[HTML]{FFCCCB}0.737 &  \cellcolor[HTML]{FFCCCB}2.342\\\hline

\multicolumn{14}{|c|}{\cellcolor[HTML]{CCECEB}Combined over and underexposed images (5,905 images)}\\ \hline
HE \cite{10.5555/559707} $\textasteriskcentered$& 16.148 & \cellcolor[HTML]{FFCCCB}0.685 & 16.283 & 0.671 & 16.525 & \cellcolor[HTML]{FFCCCB}0.696  & 16.639 & 0.668 & 17.305 & 0.688 & 16.580 & 0.682 & 2.405\\
CLAHE \cite{adaptivehisteq} $\textasteriskcentered$& 14.884 & 0.589 & 15.669 &  0.610 & 15.383 & 0.599 & 15.452 & 0.601 & 15.737 & 0.610 & 15.425 & 0.602 & 2.317\\
WVM \cite{fu2016weighted} $\textasteriskcentered$&  14.488 & 0.665 & 15.803 & \cellcolor[HTML]{FFCCCB}0.699 & 15.117 & 0.678 & 15.863 & \cellcolor[HTML]{FFCCCB}0.693 & 16.469 & \cellcolor[HTML]{FFCCCB}0.704 & 15.548 & 0.688 & 2.415\\
\cellcolor[HTML]{D5D5D5}LIME \cite{guo2016lime, guo2017lime} & 11.154 & 0.591 & 11.828 & 0.610 & 11.517 & 0.607 & 12.638 & 0.628 & 13.613 & 0.653 & 12.150 & 0.618 & 2.432\\
HDR CNN \cite{HDRCNN} w/ RHT \cite{yang2018image}& 13.709 & 0.467 & 13.921 & 0.458 & 13.800 & 0.474 & 13.716 & 0.446 & 13.558 & 0.454 & 13.741 & 0.460 & 4.599\\
HDR CNN \cite{HDRCNN} w/ PS \cite{dayley2010photoshop}  & 15.812 & 0.667 & 16.970 & 0.699 & 16.428 & 0.681 & 17.301 & 0.687 & 18.650  & 0.702 & 17.032 & \cellcolor[HTML]{FFCCCB}0.687 & \cellcolor[HTML]{FFCCCB}2.267\\
DPED (iPhone) \cite{DPED}& 15.134 & 0.609 & 16.505 & 0.636 & 15.907 & 0.622 & 16.571 & 0.627 & 17.251 & 0.649 & 16.274 & 0.629 & 2.903\\
DPED (BlackBerry) \cite{DPED} & 16.910 & 0.642 & \cellcolor[HTML]{FFCCCB}18.649 & \cellcolor[HTML]{FFCCCB}0.713 & 17.606 & 0.653 & \cellcolor[HTML]{FFCCCB}18.070 & 0.679 & 18.217 & 0.668 & \cellcolor[HTML]{FFCCCB}17.890 & 0.671 & 2.564\\
DPED (Sony) \cite{DPED}& \cellcolor[HTML]{FFCCCB}17.419  & 0.675 & 18.636 & 0.701 & \cellcolor[HTML]{FFCCCB}18.020  & 0.683 &  17.554 & 0.660 & \cellcolor[HTML]{FFCCCB}17.778 & 0.663 & 17.881 & 0.676 & 2.806\\
DPE (HDR) \cite{DPE} & 15.690 & 0.614 & 16.548 & 0.626 & 16.305 & 0.626 & 16.147 & 0.615 & 16.341 & 0.633 & 16.206 & 0.623 & 2.417\\
DPE (U-FiveK) \cite{DPE} & 16.240 & 0.653 & 16.805 & 0.646 & 16.837 & 0.671 &  16.762 & 0.654 & 16.707 & 0.650 & 16.670 & 0.655 & 2.606\\
DPE (S-FiveK) \cite{DPE} & 16.933 & 0.678 & 17.701 & 0.668 & 17.741 & \cellcolor[HTML]{FFCCCB}0.696 & 17.572 & 0.674 & 17.601 & 0.670 & 17.510 & 0.677 & 2.621\\
\cellcolor[HTML]{D5D5D5}HQEC \cite{HQEC} $\textasteriskcentered$& 13.385 & 0.641 & 14.470 & 0.666 & 13.911 & 0.656 & 14.891 & 0.674 & 15.777 & 0.692 & 14.487 & 0.666 & 2.445\\
\cellcolor[HTML]{D5D5D5}RetinexNet \cite{Chen2018Retinex}  & 10.759 & 0.585 & 11.613 & 0.596 & 11.135 & 0.605 & 11.987 & 0.615 & 12.671 & 0.636 & 11.633 & 0.607 & 3.105\\
\cellcolor[HTML]{D5D5D5}Deep UPE \cite{DeepUPE} & 13.161 & 0.610 & 13.901 & 0.642 & 13.689 & 0.632 & 14.806 & 0.649 & 15.678 & 0.667 & 14.247 & 0.640  & 2.405\\ 
\cellcolor[HTML]{D5D5D5}Zero-DCE \cite{guo2020zero}  & 11.643  & 0.536 & 12.555 & 0.539 & 12.058 & 0.544 & 12.964  & 0.548 & 13.769 & 0.580 & 12.5978 & 0.5494 & 2.865 \\
\hdashline
Our method w/o $\mathcal{L}_{\text{adv}}$ & \cellcolor[HTML]{79CC7A}\textbf{19.158} & \cellcolor[HTML]{79CC7A}\textbf{0.746} &  \cellcolor[HTML]{79CC7A}\textbf{20.096} & \cellcolor[HTML]{79CC7A}\textbf{0.734} & \cellcolor[HTML]{79CC7A}\textbf{20.205} & \cellcolor[HTML]{79CC7A}\textbf{0.769} & \cellcolor[HTML]{79CC7A}\textbf{18.975} & \cellcolor[HTML]{79CC7A}\textbf{0.719} & \cellcolor[HTML]{79CC7A}\textbf{18.983} & \cellcolor[HTML]{79CC7A}\textbf{0.727} & \cellcolor[HTML]{79CC7A}\textbf{19.483} & \cellcolor[HTML]{79CC7A}\textbf{0.739} & \cellcolor[HTML]{FFFBA3}2.251\\
Our method w/ $\mathcal{L}_{\text{adv}}$ & \cellcolor[HTML]{FFFBA3}19.114 & \cellcolor[HTML]{FFFBA3}0.743 & \cellcolor[HTML]{FFFBA3}19.960 & \cellcolor[HTML]{FFFBA3}0.723 & \cellcolor[HTML]{FFFBA3}20.080 & \cellcolor[HTML]{FFFBA3}0.763 & \cellcolor[HTML]{FFFBA3}18.868 & \cellcolor[HTML]{FFFBA3}0.709 & \cellcolor[HTML]{FFFBA3}18.864 & \cellcolor[HTML]{FFFBA3}0.719 & \cellcolor[HTML]{FFFBA3}19.377 & \cellcolor[HTML]{FFFBA3}0.731 & \cellcolor[HTML]{79CC7A}\textbf{2.247} \\\hline
\end{tabular}
}\end{center}
\vspace{-6mm}
\end{table*}

For the pixel-wise error metrics -- namely, PSNR and SSIM -- we compare the results not only against the properly exposed rendered images by Expert C but also with \textit{all} five expert photographers in the MIT-Adobe FiveK dataset \cite{fivek}. Though the expert photographers may render the same image in different ways due to differences in the camera-based rendering settings (e.g., white balance and tone mapping), a common characteristic over all rendered images by the expert photographers is that they all have fairly proper exposure settings \cite{fivek} (see Fig.~\ref{fig:experts}). For this reason, we evaluate our method against the \textit{five} expert rendered images as they all represent satisfactory exposed reference images.

\begin{figure}[t]
\centering
\includegraphics[width=\linewidth]{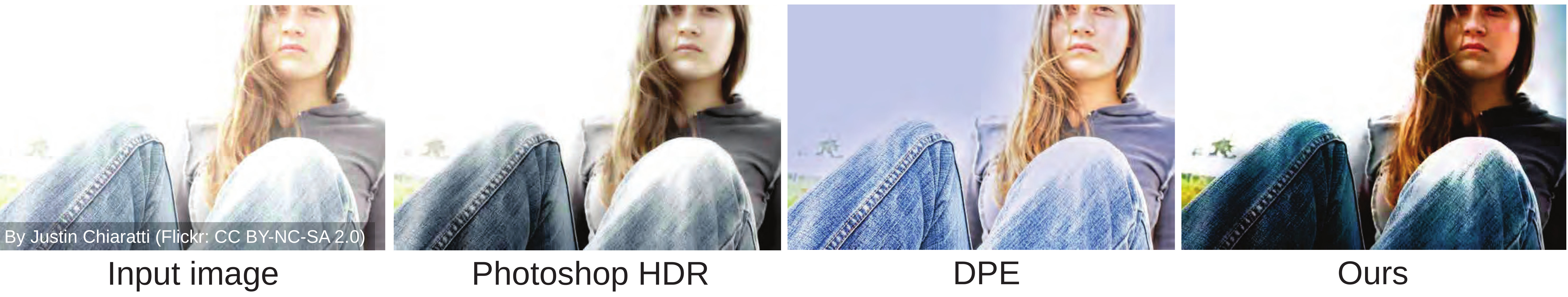}
\vspace{-6mm}
\caption{Qualitative comparison with Adobe Photoshop's local adaptation HDR function \cite{dayley2010photoshop} and DPE \cite{DPE}. Input images are taken from Flickr.\vspace{-4mm}} 
\label{fig:ours_vs_commercial_sw}
\end{figure}

We also evaluate a variety of previous non-learning and learning-based methods on our test set for comparison: histogram equalization (HE) \cite{10.5555/559707}, contrast-limited adaptive histogram equalization (CLAHE) \cite{adaptivehisteq}, the weighted variational model (WVM) \cite{fu2016weighted}, the low-light image enhancement method (LIME) \cite{guo2016lime, guo2017lime}, HDR CNN \cite{HDRCNN},  DPED models \cite{DPED},  deep photo enhancer (DPE) models \cite{DPE}, the high-quality exposure correction method (HQEC) \cite{HQEC}, RetinexNet \cite{Chen2018Retinex}, deep underexposed photo enhancer (UPE) \cite{DeepUPE}, and the zero-reference deep curve estimation method (Zero-DCE) \cite{guo2020zero}. To render the reconstructed HDR images generated by the HDR CNN method \cite{HDRCNN} back into LDR, we tested both the deep reciprocating HDR transformation method (RHT) \cite{yang2018image}, and Adobe Photoshop's (PS) HDR tool \cite{dayley2010photoshop}.

\begin{figure}
\centering
\includegraphics[width=\linewidth]{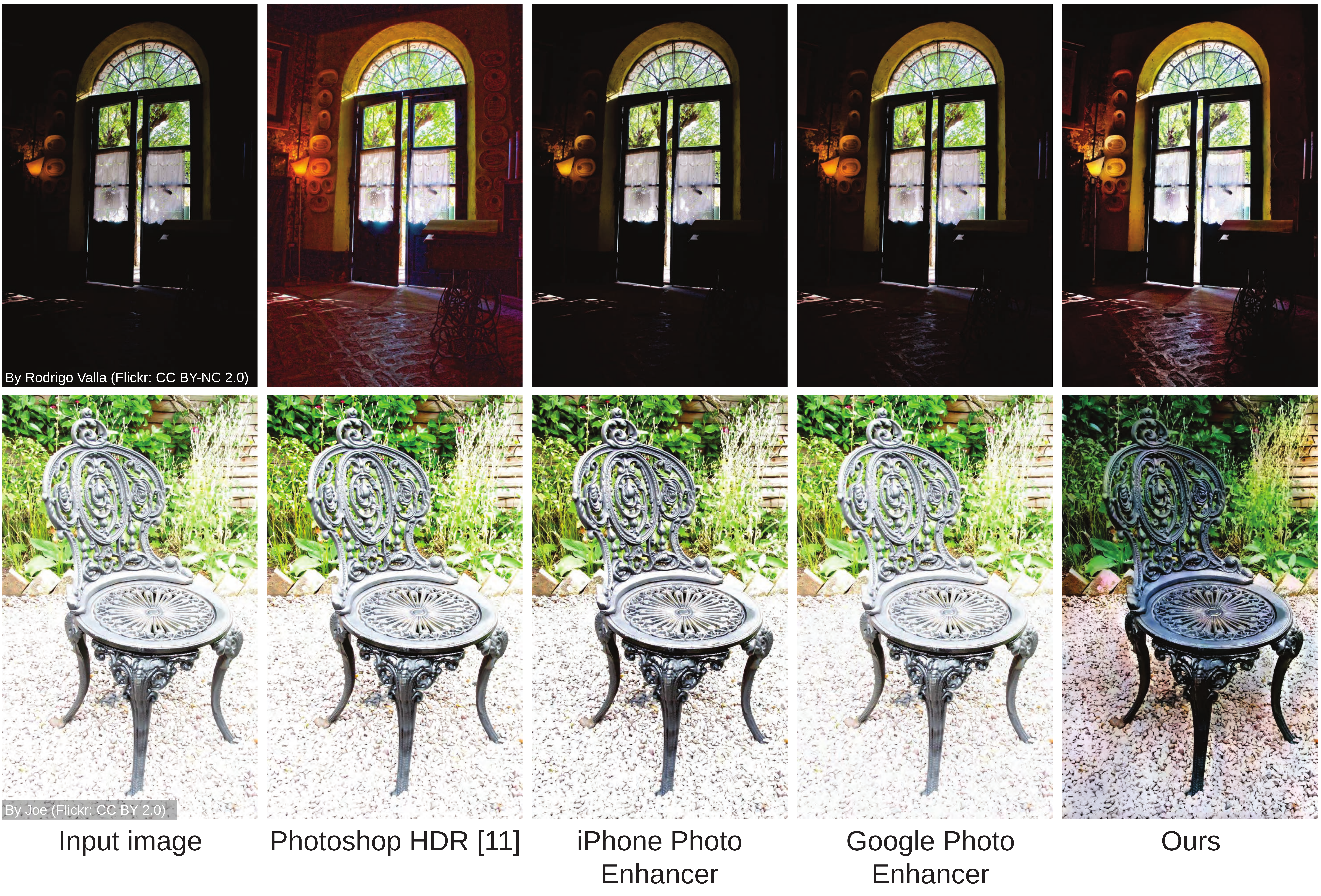}
\vspace{-6mm}
\caption{Comparisons with commercial software packages. 
The input images are taken from Flickr.\vspace{-6mm}}
\label{fig:ours_vs_commercial_sw_supp}
\end{figure}

Table\ \ref{table:results} summarizes the quantitative results obtained by each method. As shown in the top portion of the table, our method achieves the best results for overexposed images under
all metrics. In the underexposed image correction setting,
our results (middle portion of table) are on par with the state-of-the-art methods. Finally, in contrast to most of the existing methods, the results in the bottom portion of the table show that our method can effectively deal with \textit{both} types of exposure errors.

\noindent\textbf{Generalization} We further evaluate the generalization ability of our method on the following standard image datasets used by previous low-light image enhancement methods: (i)  LIME (10 images) \cite{guo2017lime}, (ii)  NPE (75 images) \cite{wang2013naturalness}, (iii)  VV (24 images) \cite{VVDataset}, and  DICM (44 images) \cite{lee2012contrast}. Note that in these experiments, we report results of our model trained on our training set without further tuning or re-training on any of these datasets. Similar to previous methods, we use the $\texttt{NIQE}$ perceptual score \cite{mittal2012making} for evaluation. Table\ \ref{table:results_extra_datasets} compares results by our method and the following methods: LIME \cite{guo2016lime, guo2017lime}, WVM \cite{fu2016weighted}, RetinexNet (RNet) \cite{Chen2018Retinex}, ``kindling the darkness'' (KinD) \cite{zhang2019kindling}, enlighten GAN (EGAN) \cite{jiang2019enlightengan}, and deep bright-channel prior (BCP) \cite{8955834}. As can be seen in Table\ \ref{table:results_extra_datasets}, our method generally achieves perceptually superior results in correcting low-light 8-bit images of other datasets.

\subsection{Qualitative Results}\label{subsec:qual-results}

We compare our method qualitatively with a variety of previous methods. Note we show results using the model trained with the adversarial loss term, as it produces perceptually superior results (see the perceptual metric results in Tables \ref{table:results} and \ref{table:results_extra_datasets}).

Fig.~\ref{fig:qualitative_our_set} shows our results on different overexposed and underexposed images. As shown, our results are arguably visually superior to the other methods, even when input images have hard backlight conditions, as shown in the second row in Fig.~\ref{fig:qualitative_our_set} (right).

\noindent\textbf{Generalization} We also ran our model on several images from Flickr that are outside our introduced dataset, as shown in Figs.\ \ref{fig:teaser}, \ref{fig:ours_vs_commercial_sw}, and  \ref{fig:ours_vs_commercial_sw_supp}.  As with the images from our introduced dataset, our results on the Flickr images are arguably superior to the compared methods. Additional qualitative results and comparisons are provided in the supplementary materials.

\subsection{Limitations}\label{subsec:limitations}
\begin{figure}[b]
\centering
\includegraphics[width=\linewidth]{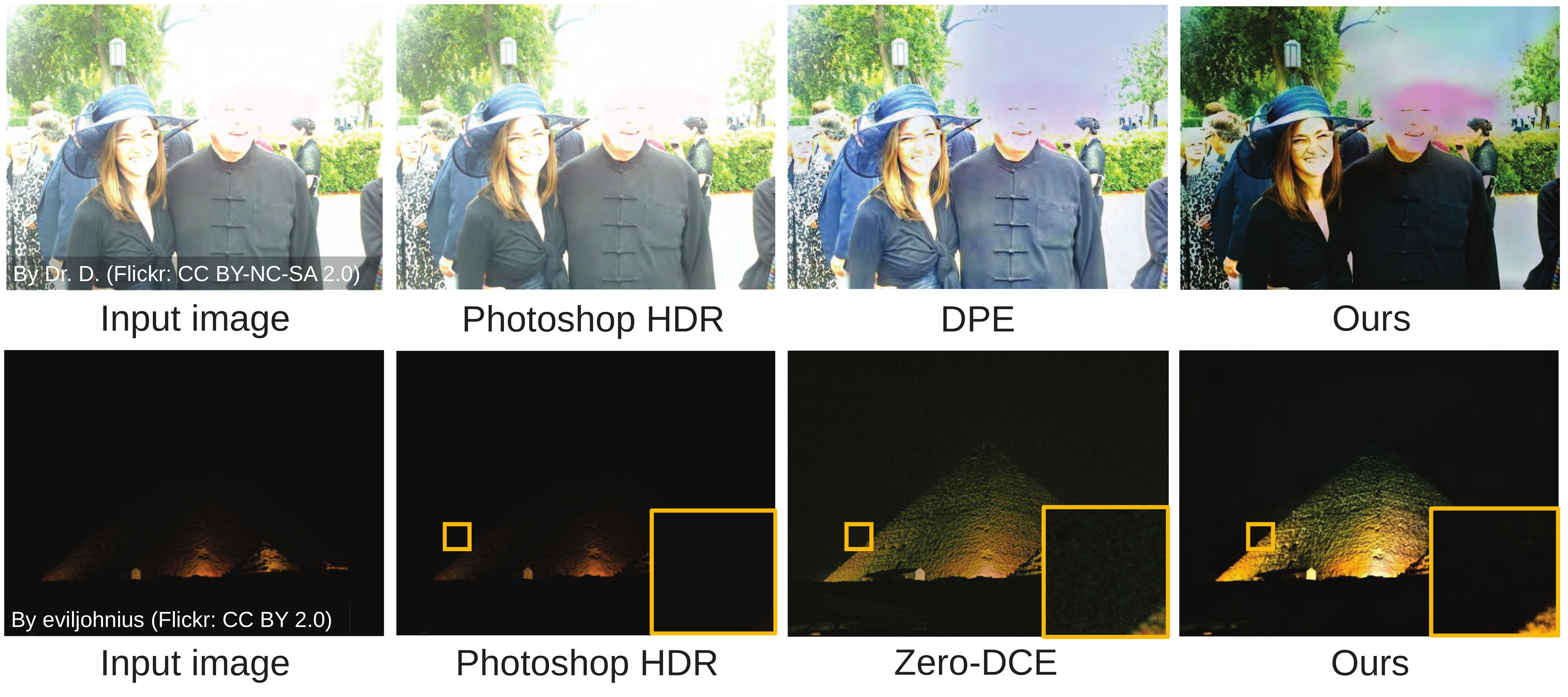}
\vspace{-6mm}
\caption{Failure examples of correcting (top) overexposed and (bottom) underexposed images. The input images are taken from Flickr.\vspace{-2mm}} 
\label{fig:failureExamples}
\end{figure}

Our method produces unsatisfactory results in regions that have insufficient semantic information, as shown in Fig.~\ref{fig:failureExamples}. For example, the input image shown in the first row in Fig.\ \ref{fig:failureExamples} is completely saturated and contains almost no details in the region of the man's face. We can see that our network cannot constrain the color inside the face region due to the lack of semantic information. In the supplementary materials, we provide a way to interactively control the output results by scaling each layer of the Laplacian pyramid before feeding them to the network. In that way, one can control the output results to reduce such color bleeding problems. It also can be observed that our method may introduce noise when the input image has extreme dark regions, as shown in the second example in Fig.\ \ref{fig:failureExamples}. These challenging conditions prove difficult for other methods as well. 

\begin{table}[t]
\caption{
Perceptual quality evaluation.  Summary of $\texttt{NIQE}$ scores \cite{mittal2012making} on different \textit{low-light} image datasets. In these dataset, there are no ground-truth images provided for full-reference quality metrics (e.g., PSNR). Highlights are in the same format as Table\ \ref{table:results}.}
\vspace{-4mm}
\label{table:results_extra_datasets}
\begin{center}
\scalebox{0.67}{
\begin{tabular}{|l|c|c|c|c|c|}
\hline
Method & LIME \cite{guo2017lime} & NPE \cite{wang2013naturalness} & VV \cite{VVDataset} & DICM \cite{lee2012contrast} & Avg. \\ \hline
NPE \cite{wang2013naturalness} $\textasteriskcentered$ & 3.91 & 3.95 & \cellcolor[HTML]{FFCCCB}2.52 & 3.76 & 3.54 \\ \hline
LIME \cite{guo2017lime} $\textasteriskcentered$ & 4.16 & 4.26 & \cellcolor[HTML]{FFFBA3}2.49 & 3.85 & 3.69 \\ \hline
WVM \cite{fu2016weighted} $\textasteriskcentered$ & 3.79 & 3.99 & 2.85 & 3.90 & 3.63 \\ \hline
RNet \cite{Chen2018Retinex} & 4.42 & 4.49 & 2.60 & 4.20 & 3.93 \\ \hline
KinD \cite{zhang2019kindling} & \cellcolor[HTML]{79CC7A}\textbf{3.72} & \cellcolor[HTML]{FFCCCB}3.88 & - & - & 3.80 \\ \hline
EGAN \cite{jiang2019enlightengan} & \cellcolor[HTML]{79CC7A}\textbf{3.72} & 4.11 & 2.58 & - & 3.50 \\ \hline
DBCP \cite{8955834} & \cellcolor[HTML]{FFCCCB}3.78 & \cellcolor[HTML]{79CC7A}\textbf{3.18} & - & \cellcolor[HTML]{FFCCCB}3.57 & \cellcolor[HTML]{FFCCCB}3.48 \\ \hdashline
Ours w/o $\mathcal{L}_{\text{adv}}$ & \cellcolor[HTML]{FFFBA3}3.76 & \cellcolor[HTML]{FFFBA3}3.20 & \cellcolor[HTML]{79CC7A}\textbf{2.28} & \cellcolor[HTML]{FFFBA3}2.55 & \cellcolor[HTML]{FFFBA3}2.95 \\ \hline
Ours w/ $\mathcal{L}_{\text{adv}}$ & \cellcolor[HTML]{FFFBA3}3.76 & \cellcolor[HTML]{79CC7A}\textbf{3.18} & \cellcolor[HTML]{79CC7A}\textbf{2.28} & \cellcolor[HTML]{79CC7A}\textbf{2.50} & \cellcolor[HTML]{79CC7A}\textbf{2.93} \\ \hline
\end{tabular}
}\end{center}
\vspace{-6mm}
\end{table}

\section{Concluding Remarks}\label{sec:conclusion}

We proposed a single coarse-to-fine deep learning model for overexposed and underexposed image correction. We employed the Laplacian pyramid decomposition to process input images in different frequency bands. Our method is designed to sequentially correct each of the Laplacian pyramid levels in a multi-scale manner, starting with the global color in the image and progressively addressing the image details. 

Our method is enabled by a new dataset of over 24,000 images rendered with the broadest range of exposure errors to date. Each image in our introduced dataset has a reference image properly rendered by a well-trained photographer with well-exposure compensation. Through extensive evaluation, we showed that our method produces compelling results compared to available solutions for correcting images rendered with exposure errors and it generalizes well. We believe that our dataset will help future work on improving exposure correction for photographs.

\newcommand{\beginsupplement}{%
        \setcounter{table}{0}
        \renewcommand{\thetable}{S\arabic{table}}%
        \setcounter{figure}{0}
        \renewcommand{\thefigure}{S\arabic{figure}}%
     }

\section{Supplementary Material}
\beginsupplement

\begin{figure*}[]
\centering
\vspace{-1mm}
\includegraphics[width=\textwidth]{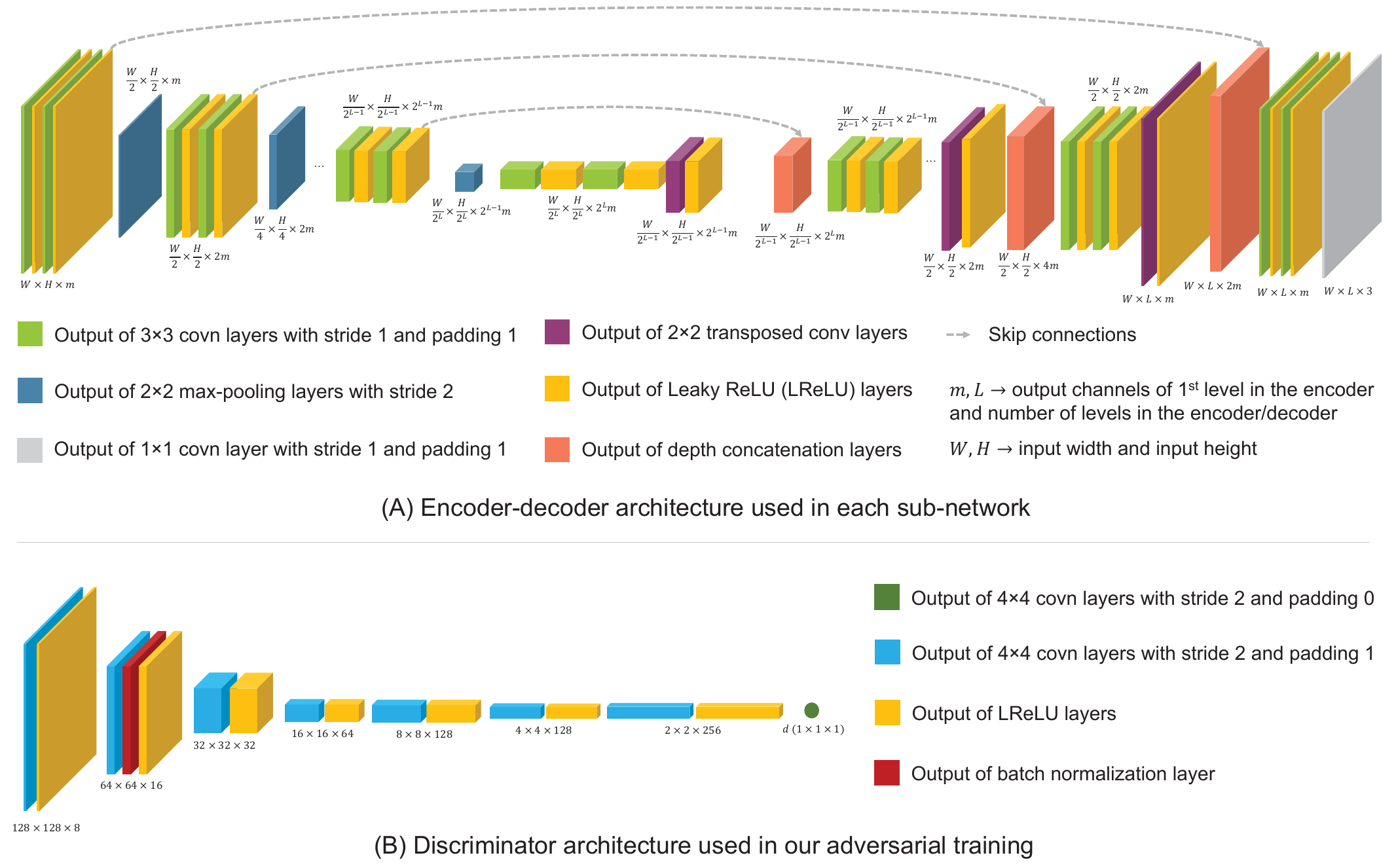}
\vspace{-6mm}
\captionof{figure}{Details of the architectures used in our work. (A) Encoder-decoder architecture \cite{unet} used to design our sub-networks in the main network. (B) Discriminator architecture.\vspace{13mm}} \label{fig:arch}
\end{figure*}

\begin{figure*}[]
\centering
\includegraphics[width=0.85\linewidth]{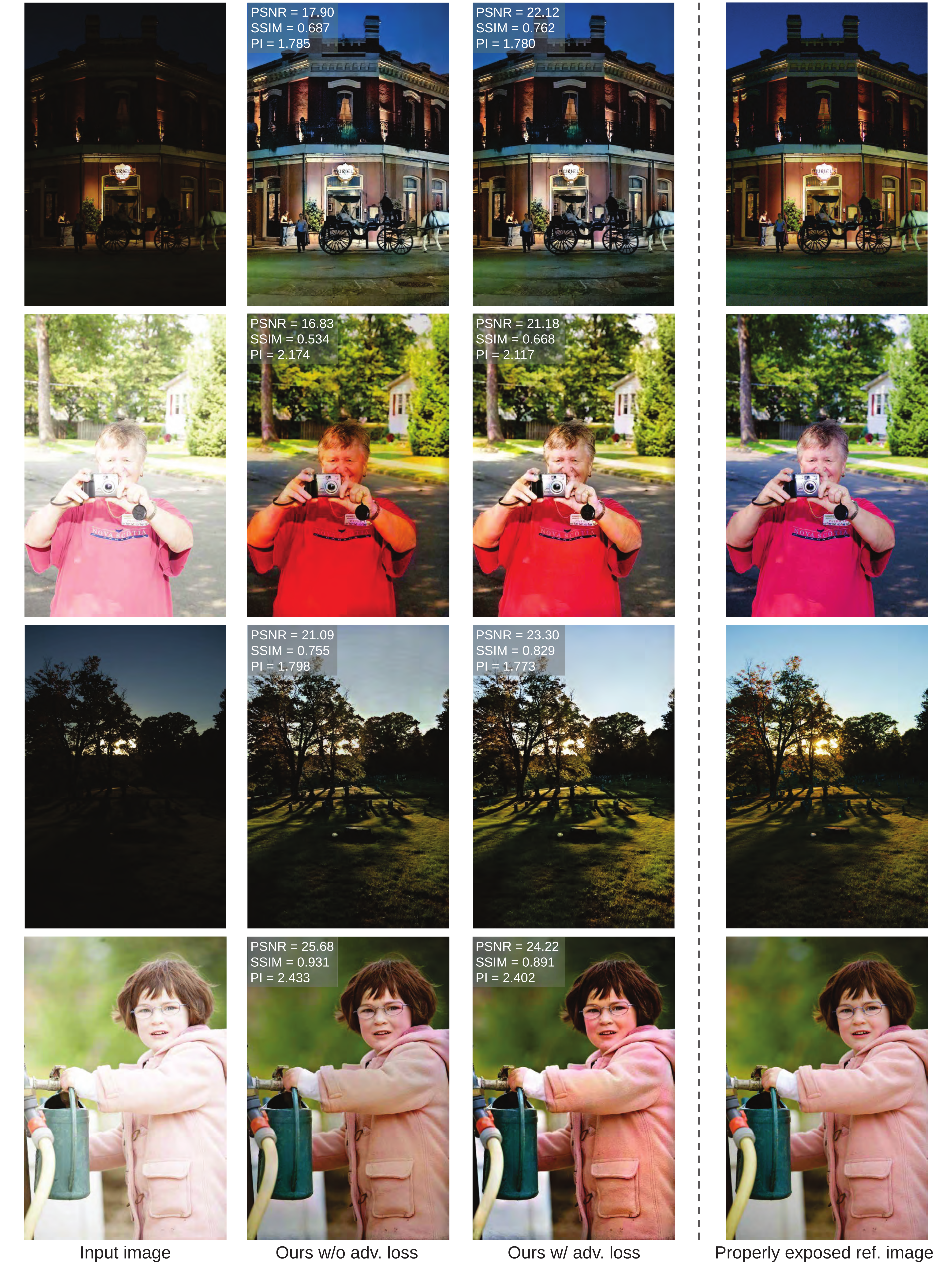}
\vspace{-2mm}
\caption{Comparisons between our results with (w/) and without (w/o) the adversarial loss for training. The peak signal-to-noise ratio (PSNR), structural similarity index measure (SSIM) \cite{wang2004image}, and perceptual index (PI) \cite{blau20182018} are shown for each result. Notice that higher PSNR and SSIM values are better, while lower PI values indicate better perceptual quality. The input images are taken from our test set.\vspace{-2mm}} \label{fig:ablation-with_without_discriminator}
\end{figure*}

\subsection{Implementation Details}\label{sec:implementation_details}

In the main paper, we proposed a coarse-to-fine network to correct exposure errors in photographs. In this section, we provide the implementation details of our network, the discriminator network used in the adversarial training process, and additional training details. 

\subsubsection{Main Network}

\begin{figure*}[t]
\centering
\includegraphics[width=\linewidth]{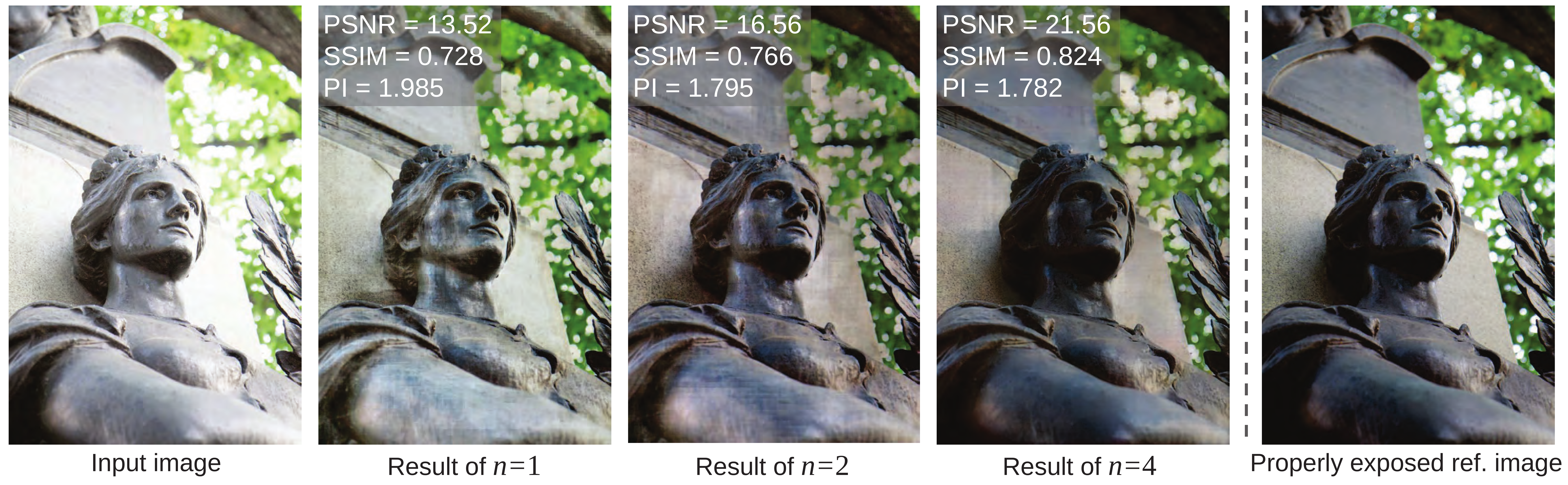}
\vspace{-7mm}
\caption{Comparison of results by varying the number of Laplacian pyramid levels. 
The peak signal-to-noise ratio (PSNR), structural similarity index measure (SSIM) \cite{wang2004image}, and perceptual index (PI) \cite{blau20182018} are shown for each result. Notice that higher PSNR and SSIM values are better, while lower PI values indicate better perceptual quality. The input image is taken from our validation set.\vspace{-2mm}}
\label{fig:ablation1}
\end{figure*}

Our main network consists of four sub-networks with $\sim$7M parameters trained in an end-to-end manner. The largest network capacity is dedicated to the first sub-network with decreasing amounts of capacity as we move from coarse-to-fine scales. Each sub-network accepts a different representation of the input image extracted from the Laplacian pyramid decomposition. The first sub-network is a four-layer encoder-decoder network with skip connections (i.e., U-Net-like architecture  \cite{unet}). The output of the first convolutional (conv) layer has 24 channels. Our first sub-network has $\sim$4.4M learnable parameters and accepts the low-frequency band level of the Laplacian pyramid, i.e., $\mathbf{X}_{(4)}$. The result of the first sub-network is then upscaled using a $2\!\times\!2\!\times\!3$ transposed conv layer with three output channels and a stride of two. This processed layer is then added to the first mid-frequency band level of the Laplacian pyramid (i.e., $\mathbf{X}_{(3)}$) and is fed to the second sub-network. 

The second sub-network is a three-layer encoder-decoder network with skip connections. It has 24 channels in the first conv layer of the encoder, with a total of $\sim$1.1M learnable parameters. The second sub-network processes the upscaled input from the first sub-network and outputs a residual layer, which is then added back to the input to the second sub-network followed by a $2\!\times\!2\!\times\!3$ transposed conv layer with three output channels and a stride of two. The result is added to the second mid-frequency band level of the Laplacian pyramid (i.e., $\mathbf{X}_{(2)}$) and is fed to the third sub-network, which generates a new residual that is added back again to the input of this sub-network. 

The third sub-network has the same design as the second network.  Finally, the result is added to the high-frequency band level of the Laplacian pyramid (i.e., $\mathbf{X}_{(1)}$) and is fed to the fourth sub-network to produce the final processed image.

The final sub-network is a three-layer encoder-decoder network with skip connections and has $\sim$482.2K learnable parameters, where the output of the first conv layer in its encoder has 16 channels. We provide the details of the main encoder-decoder architecture of each sub-network in Fig.\ \ref{fig:arch}-(A).

\subsubsection{Discriminator Network}
In the adversarial training of our network, we use a light-weight discriminator network with $\sim$1M learnable parameters. We provide the details of the discriminator in Fig.\ \ref{fig:arch}-(B). Notice that unlike our main network, we resize all input image patches to have $256\!\times\!256$ pixels before being processed by the discriminator. The output of the last layer in our discriminator is a single scalar value which is then used in our loss during the optimization, as described in the main paper.

 \begin{figure*}[t]
 \centering
 \includegraphics[width=0.95\linewidth]{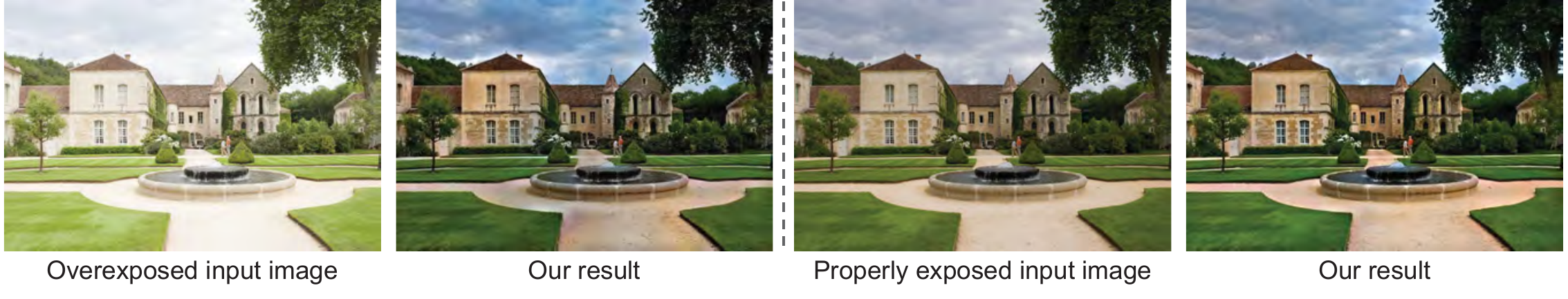}
 \vspace{-2mm}
 \caption{Our framework can deal with both improperly and properly exposed input images producing compelling results. The input images are taken from our test set.\vspace{-2mm}}
 \label{fig:input_well_exposed}
 \end{figure*}

\begin{figure*}[t]
\centering
  \includegraphics[width=0.9\linewidth]{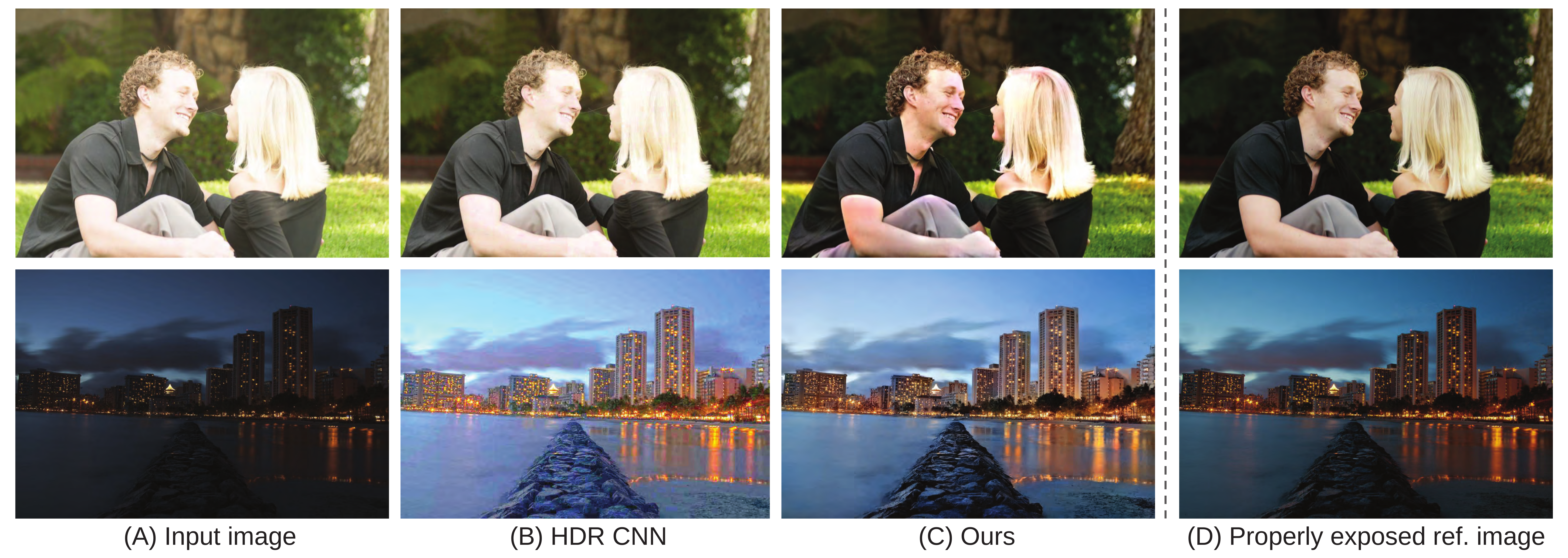}
  \vspace{-2mm}
  \caption{Additional qualitative results. (A) Input images. (B) Results of HDR CNN \cite{HDRCNN} with Adobe Photoshop's HDR tool \cite{dayley2010photoshop}. (C) Our results. (G) Properly exposed reference images. The input images are taken from our test set.\vspace{-2mm}}
  \label{fig:qualitative_comparison_with_HDRCNN}
  \end{figure*}

\begin{figure*}[t]
 \centering
 \includegraphics[width=0.92\linewidth]{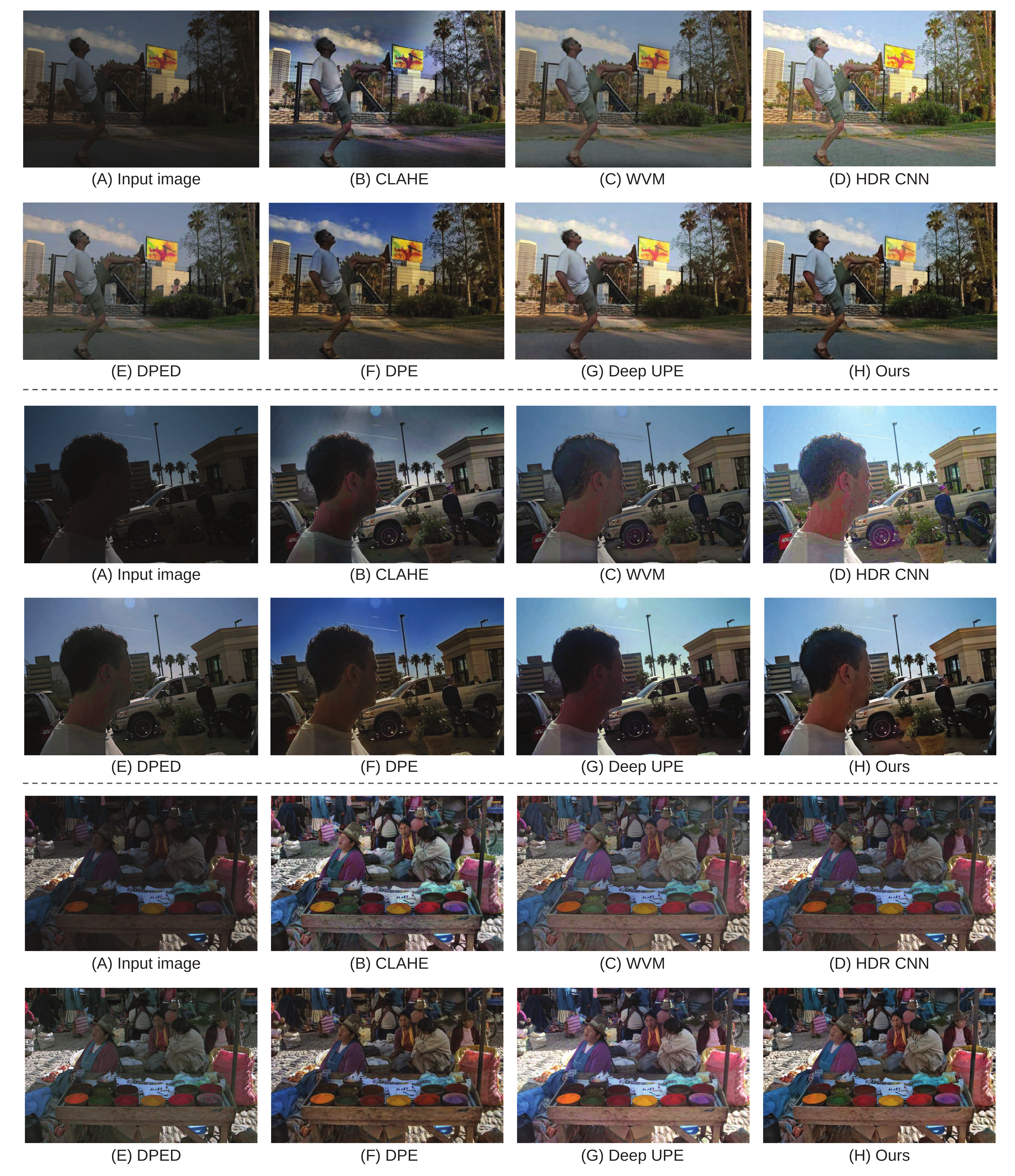}
 \vspace{-2mm}
 \caption{Additional qualitative comparisons with other methods in correcting underexposed images. (A) Input images. (B) Results of CLAHE \cite{adaptivehisteq}. (C) Results of WVM \cite{fu2016weighted}. (D) Results of HDR CNN \cite{HDRCNN} with Adobe Photoshop's HDR tool \cite{dayley2010photoshop}. (E) Results of DPED \cite{DPED}. (F) Results of DPE \cite{DPE}. (G) Results of Deep UPE \cite{DeepUPE}. (H) Our results. The input images are taken from our test set. \vspace{-2mm}}
 \label{fig:qualitative_other_methods_underexp}
 \end{figure*}
 
  \begin{figure*}[t]
  \centering
  \includegraphics[width=0.9\linewidth]{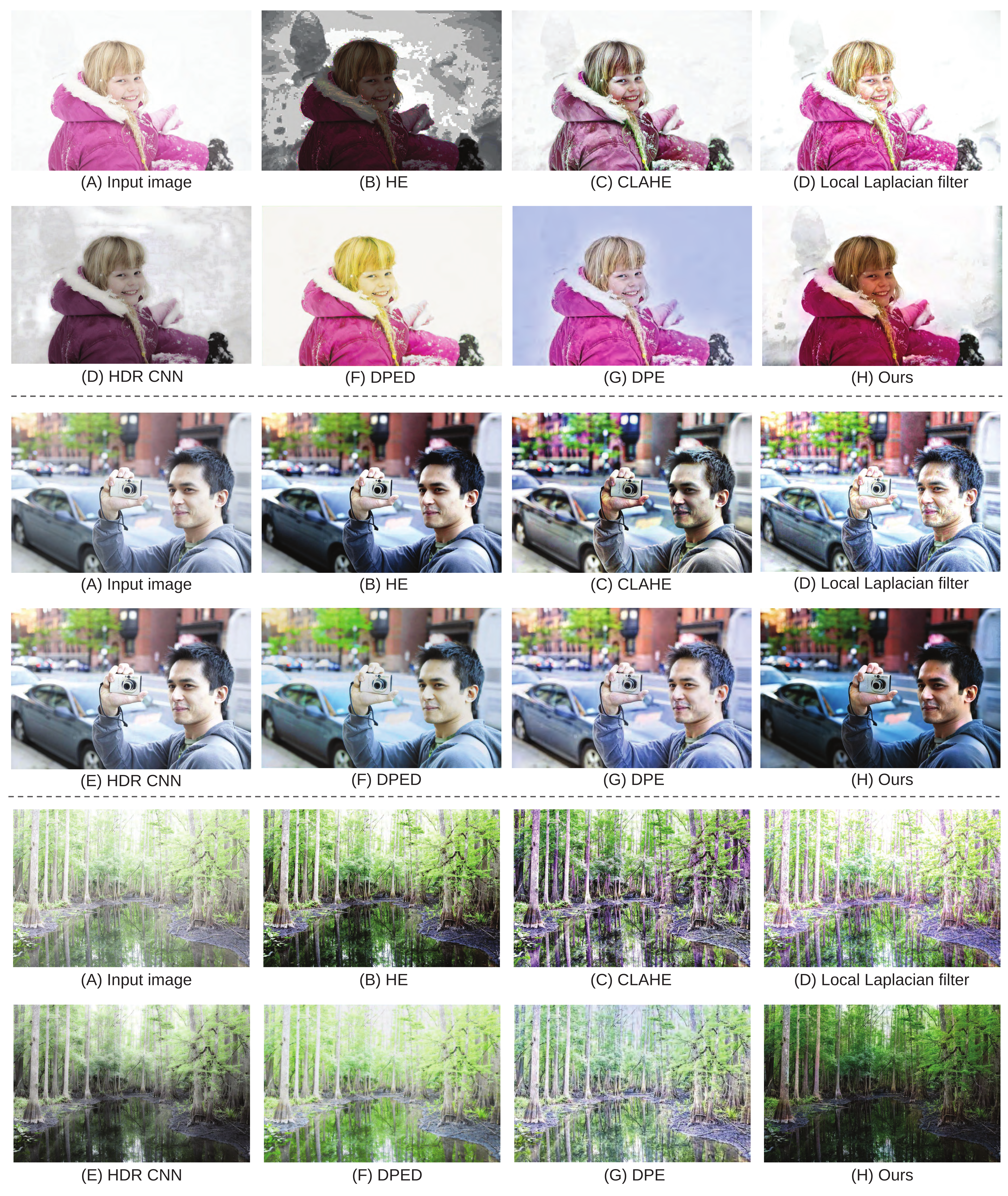}
  \vspace{-2mm}
  \caption{Additional qualitative comparisons with other methods in correcting overexposed images. (A) Input images. (B) Results of histogram equalization (HE) \cite{10.5555/559707}. (C) Results of the contrast-limited adaptive histogram equalization (CLAHE) \cite{adaptivehisteq}. (D) Results of the local Laplacian filter \cite{paris2015local}. (E) Results of HDR CNN \cite{HDRCNN} with Adobe Photoshop's (PS) HDR tool \cite{dayley2010photoshop}. (F) Results of the DSLR Photo Enhancement dataset (DPED) trained model \cite{DPED}. (G) Results of deep photo enhancer (DPE) \cite{DPE}. (H) Our results. The input images are taken from our test set.\vspace{-4mm}}
  \label{fig:qualitative_other_methods_overexp}
  \end{figure*}

 \begin{figure*}[t]
 \centering
 \includegraphics[width=0.9\linewidth]{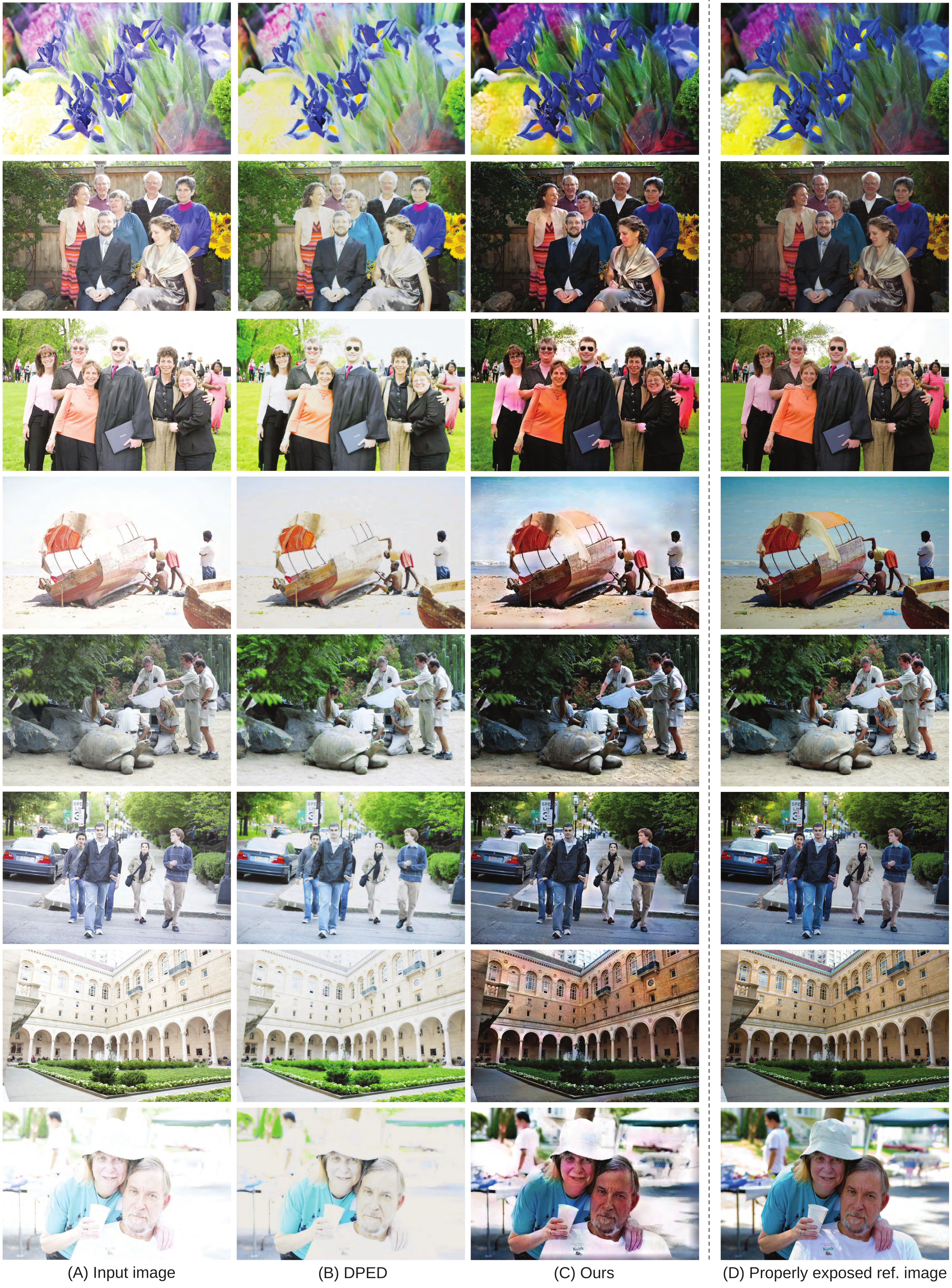}
 \vspace{-2mm}
 \caption{Additional qualitative results of correcting overexposed images. (A) Input images. (B) Results of DPED \cite{DPED}. (C) Our results. (G) Properly exposed reference images. The input images are taken from our test set.\vspace{-2mm}}
 \label{fig:qualitative_our_set_over_supp}
 \end{figure*}
 
 \begin{figure*}[t]
 \centering
 \includegraphics[width=0.9\linewidth]{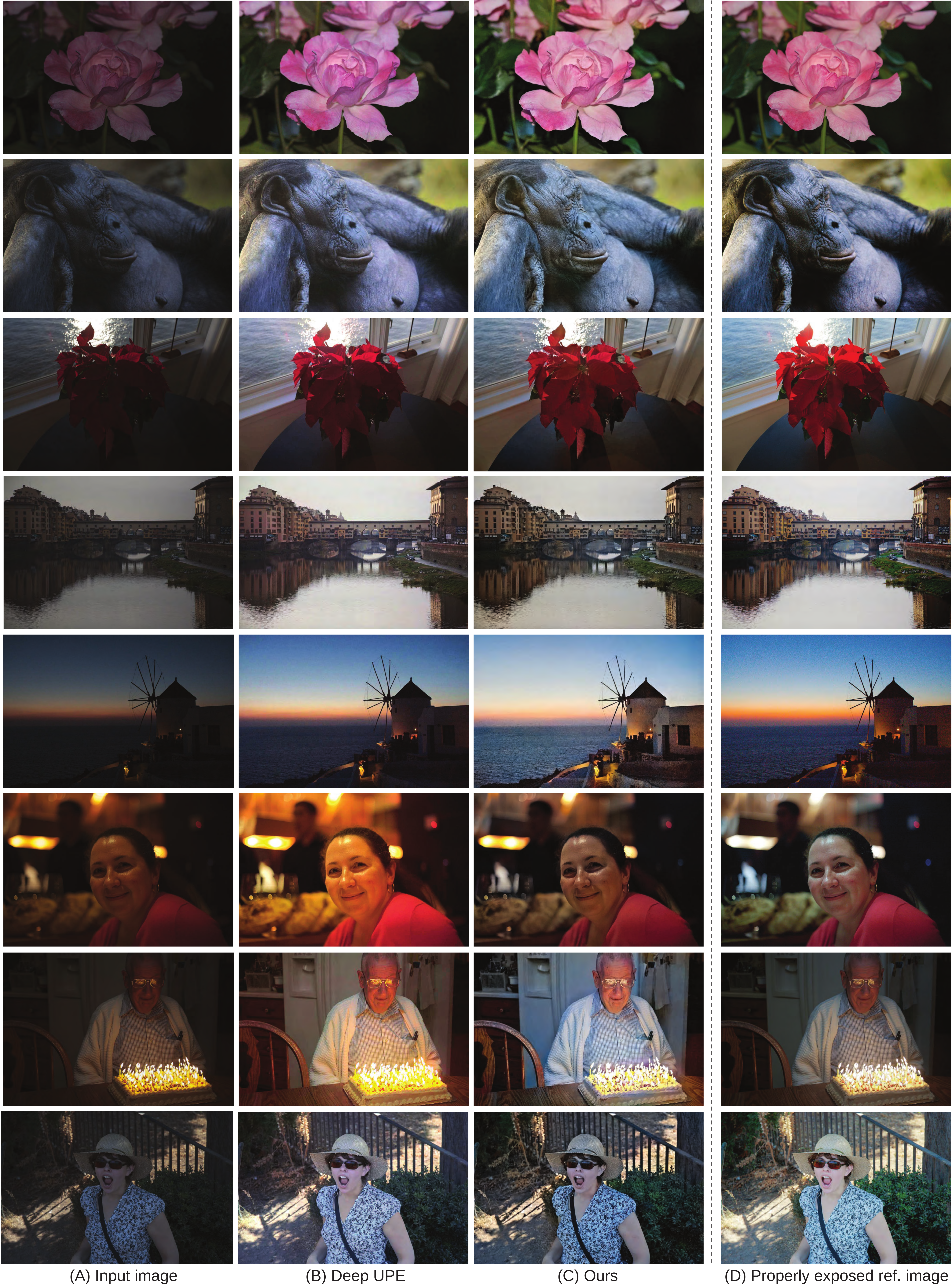}
 \vspace{-2mm}
 \caption{Additional qualitative results of correcting underexposed images. (A) Input images. (B) Results of Deep UPE \cite{DeepUPE}. (C) Our results. (G) Properly exposed reference images. The input images are taken from our test set.\vspace{-2mm}}
 \label{fig:qualitative_our_set_under_supp}
 \end{figure*}

 
 \begin{figure*}[t]
 \centering
 \includegraphics[width=0.9\linewidth]{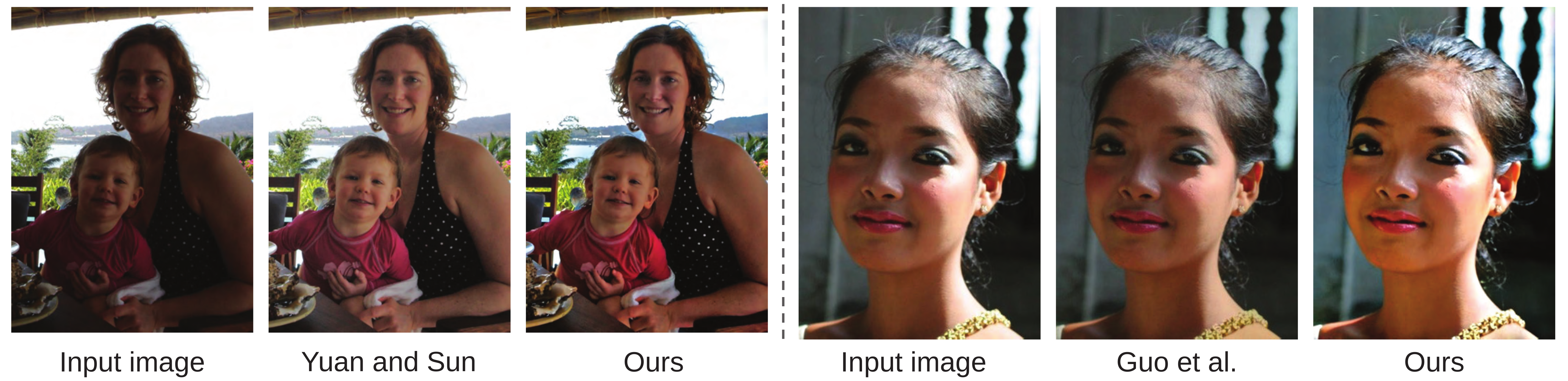}
 \vspace{-2mm}
 \caption{Qualitative comparison with the methods of Yuan and Sun \cite{yuan2012automatic} and Guo et al. \cite{guo2010correcting}. The input images are taken from  \cite{yuan2012automatic} and \cite{guo2010correcting}, respectively.}
 \label{fig:comparisons_old_methods}
 \end{figure*}

\subsubsection{Additional Training Details}\label{subsec:training_details_supp} 
We use He et al.'s method \cite{he2015delving} to initialize the weights of our encoder and decoder conv layers, while the bias terms are initialized to zero. We minimize our loss functions using the Adam optimizer \cite{kingma2014adam} with a decay rate $\beta_1 = 0.9$ for the exponential moving averages of the gradient and a decay rate $\beta_2 = 0.999$ for the squared gradient. We use a learning rate of $10^{-4}$ to update the parameters of our main network and a learning rate of $10^{-5}$ to update our discriminator's parameters.

We train our network on patches with different dimensions. Training begins without the adversarial loss, $\mathcal{L}_{\text{adv}}$, then $\mathcal{L}_{\text{adv}}$ is added to fine-tune the results of our initial training \cite{ma2017pose}. Specifically, we begin our training without $\mathcal{L}_{\text{adv}}$ on 176,590 patches with dimensions of $128\!\times\!128$ pixels extracted randomly from our training images for 40 epochs. The mini-batch size is set to 32. The learning rate is decayed by a factor of 0.5 after the first 20 epochs. Then, we continue training on another 105,845 patches with dimensions of $256\!\times\!256$ pixels for 30 epochs with a mini-batch size of eight. At this stage, we train our main network without $\mathcal{L}_{\text{adv}}$ for 15 epochs and continue training for another 15 epochs with $\mathcal{L}_{\text{adv}}$. The learning rates for the main network and the discriminator network are decayed by a factor of 0.5 every 10 epochs. Finally, we fine-tune the trained networks on another 69,515 training patches with dimensions of $512\!\times\!512$ pixels for 20 epochs with a mini-batch size of four and a learning rate decay of 0.5 applied every five epochs.

We discard any training patches that have an average intensity less than 0.02 or higher than 0.98. We also discard homogeneous patches that have an average gradient magnitude less than 0.06. We randomly left-right flip training patches for data augmentation.

In the adversarial training, we optimize both the main network and the discriminator in an iterative manner. At each optimization step, the learnable parameters of each network are updated to minimize its own loss function. Our main network's loss function is described in the main paper. The discriminator is trained to minimize the following loss function \cite{goodfellow2014generative}:

\begin{equation}
\label{eq:dsc_loss}
\mathcal{L}_{\text{dsc}} = r\left(\mathbf{T}\right) + c\left(\mathbf{Y}\right),
\end{equation}
where $r\left(\mathbf{T}\right)$ refers to the discriminator loss of recognizing the properly exposed reference image $\mathbf{T}$, while $c\left(\mathbf{Y}\right)$ refers to the discriminator loss of recognizing our corrected image $\mathbf{Y}$. The $r\left(\mathbf{T}\right)$ and $c\left(\mathbf{Y}\right)$ loss functions are given by the following equations: 
\begin{equation}
\label{eq:r_loss}
r\left(\mathbf{T}\right) =
-\log\left(\mathcal{S}\left(\mathcal{D}\left(\mathbf{T}\right)\right)\right), 
\end{equation}

\begin{equation}
\label{eq:e_loss}
c\left(\mathbf{Y}\right) = 
-\log\left(1-\mathcal{S}\left(\mathcal{D}\left(\mathbf{Y}\right)\right)\right),
\end{equation}    

\noindent where $\mathcal{S}$ denotes the sigmoid function and $\mathcal{D}$ is the discriminator network described in Fig.\ \ref{fig:arch}-(B).

\subsection{Ablation Studies (Loss Function)}\label{sec:ablation_study}

Our loss function (Eq.\ 1 in the main paper) includes three main terms. The first term is the standard reconstruction loss (i.e., $\texttt{L}_1$ loss). The second and third terms consist of the pyramid and adversarial losses, respectively, which are introduced to further improve the reconstruction and perceptual quality of the output images. In the following, we discuss the effect of these loss terms.

\begin{table}[t]
\caption{Results of our ablation study on 500 images randomly selected from our validation set. We show the effects of: (i) the pyramid loss, $\mathcal{L}_{\text{pyr}}$, and (ii) the number of Laplacian levels, $n$, in the main network. For each experiment, we show the values of the peak signal-to-noise ratio (PSNR) and structural similarity index measure (SSIM) \cite{wang2004image}. The best PSNR/SSIM values are indicated with bold for each experiment.\vspace{-2mm}}\label{table:ablation}
\scalebox{0.92}{
\begin{tabular}{l|c|c:c|c|c|}
\cline{2-6}
 & \multicolumn{2}{c:}{\cellcolor[HTML]{FFDDDA} Pyramid loss $\mathcal{L}_{\text{pyr}}$} & \multicolumn{3}{c|}{\cellcolor[HTML]{CCECEB} Number of levels $n$} \\ \cline{2-6} 
 &  w/o & w/ & $n=1$ & $n=2$ & $n=4$ \\ \hline
\multicolumn{1}{|l|}{PSNR} & 18.041 & \textbf{18.385}  & 16.984 & 17.442 & \textbf{18.385}  \\ \hline
\multicolumn{1}{|l|}{SSIM} & 0.746 & \textbf{0.749}  & 0.723 & 0.734 & \textbf{0.749}  \\ \hline
\end{tabular}\vspace{-2mm}}
\end{table}

\subsubsection{Pyramid Loss Impact} \label{subsubsec:pyramid}

In Fig.\ 5 of the main paper, we show the output of each sub-network when we train our model with and without the pyramid loss. We observe that the pyramid loss helps to provide additional supervision to guide each sub-network to follow a coarse-to-fine reconstruction. In this ablation study, we aim to quantitatively evaluate the effect of the pyramid loss on our final results.

We train two light-weight models of our main network with and without our pyramid loss term. Each model has four 3-layer U-Nets with a total of $\sim$4M learnable parameters, where the number of output channels of the first encoder in each U-Net is set to 24. 

The training is performed on a sub-set of our training data for $\sim$150,000 iterations on 80,000 $128\!\times\!128$ patches, $\sim$100,000 iterations on 40,000 $256\!\times\!256$ patches, and $\sim$25,000 iterations on 25,000 $512\!\times\!512$ patches. Table\ \ref{table:ablation} shows the results on 500 randomly selected images from our validation set. The results show that the pyramid loss not only helps in providing a better interpretation of the task of each sub-network but also improves the final results.

\subsubsection{Adversarial Loss Impact}

In the main paper, we show quantitative results of our method with and without the adversarial loss term. Our trained model with the adversarial loss term achieves better perceptual quality (i.e., lower perceptual index (PI) values \cite{blau20182018}) than training without the adversarial loss term.

Fig.\ \ref{fig:ablation-with_without_discriminator} shows qualitative comparisons of our results with and without the adversarial loss. As shown, the network trained without the adversarial training tends to produce darker images with slightly unrealistic colors in some cases, while the adversarial regularization improves the perceptual quality of our results.

\begin{figure*}[t]
\centering
\includegraphics[width=0.9\linewidth]{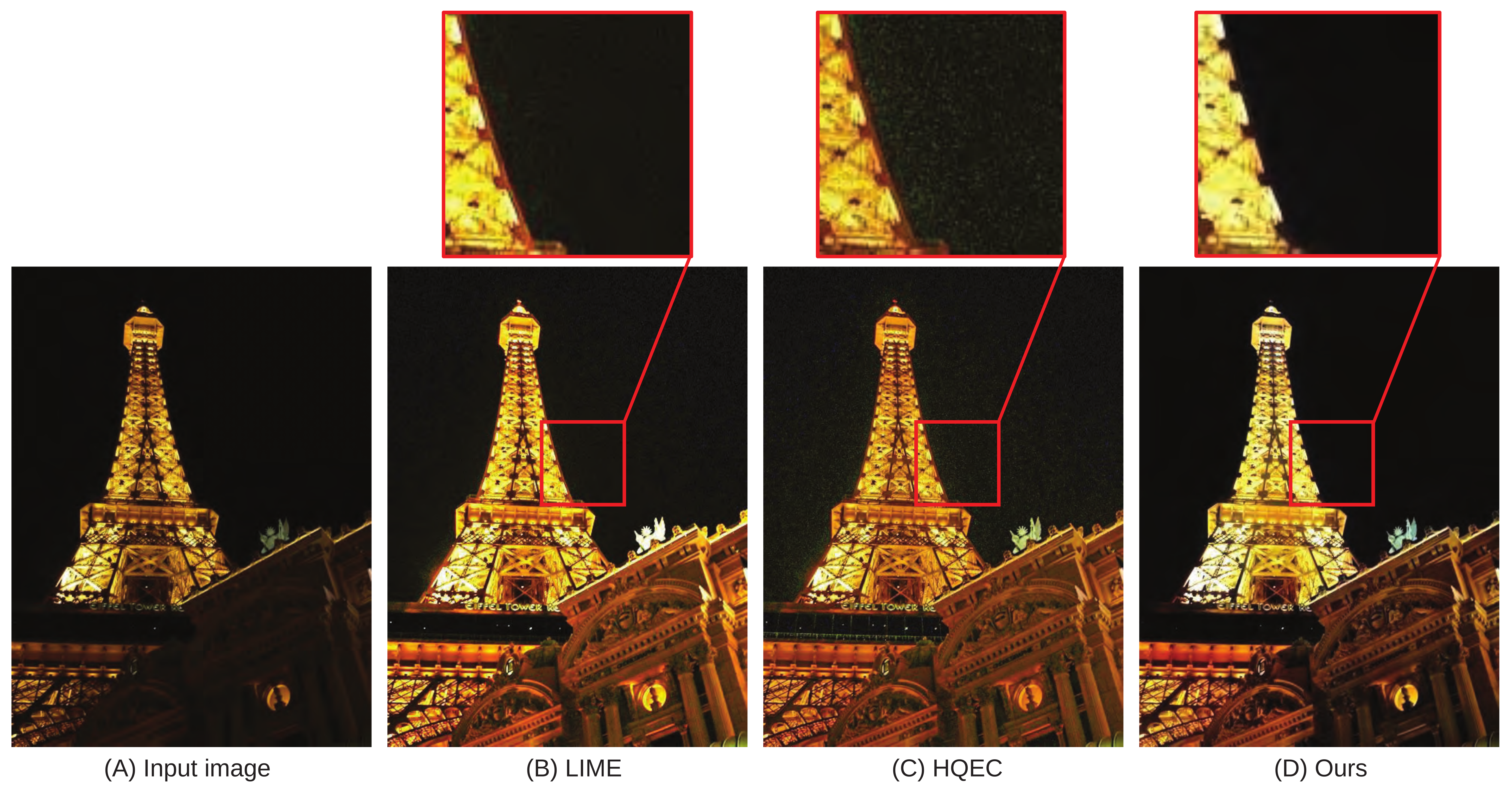}
\vspace{-2mm}
\caption{Additional qualitative results of correcting overexposed images. (A) Input image. (B) Result of LIME \cite{guo2016lime, guo2017lime}. (C) Result of HQEC \cite{HQEC}. (D) Our result. 
The input image is taken from the DICM image set \cite{lee2012contrast}.}
\label{fig:qualitative_comparisons_on_DICM}
\end{figure*}

\subsection{Ablation Studies (Number of Laplacian Pyramid Levels)}

We repeat the same experimental setup described in Sec.\ \ref{subsubsec:pyramid} with a varying number of Laplacian pyramid levels (sub-networks). Specifically, we train a network with $n=1$ levels---this network is equivalent to a vanilla U-Net-like architecture \cite{unet}. Additionally, we train another network with $n=2$ (i.e., two sub-networks).

For a fair comparison, we fix the total number of parameters in each model by changing the number of filters in the conv layers. Specifically, we set the number of output channels of the first layer in the encoder to 48 for the trained model with $n=1$, while we decrease it to 34 for the two-sub-net model (i.e., $n=2$) to have approximately the same number of learnable parameters.  Thus, the trained model in Sec.\ \ref{subsubsec:pyramid}, used to study the pyramid loss impact, and the additional two trained models have approximately the same number of parameters.

Table\ \ref{table:ablation} shows the results obtained by each model on the same random validation image subset used to study the pyramid loss impact in Sec.\ \ref{subsubsec:pyramid}. Fig.\ \ref{fig:ablation1} shows a qualitative comparison. As can be seen, the best quantitative and qualitative results are obtained using the four-sub-net model  (i.e., $n=4$ levels).

\subsection{Additional Results and Comparisons}\label{sec:additional_results}

In this section, we provide additional qualitative results. Fig.\ \ref{fig:input_well_exposed} shows our results when the input image has no exposure errors. As can be seen, our method produces consistent output images regardless of the exposure setting of the input image. Additional qualitative comparisons with other methods on our testing set are shown in Fig.\ \ref{fig:qualitative_comparison_with_HDRCNN}--\ref{fig:qualitative_our_set_under_supp}.

\paragraph{Generalization}
We provide additional results on images that are outside our training/testing sets. Fig.\ \ref{fig:comparisons_old_methods} shows qualitative comparisons with the methods of Yuan and Sun \cite{yuan2012automatic} and Guo et al. \cite{guo2010correcting}, which were designed to correct overexposure errors in photographs.  The source code of these methods is not available. Thus, the presented input images and corresponding results by the methods of Yuan and Sun \cite{yuan2012automatic} and Guo et al. \cite{guo2010correcting} are taken from the original papers \cite{yuan2012automatic, guo2010correcting}. As shown in Fig.\ \ref{fig:comparisons_old_methods}, our method produces compelling results. 

\begin{figure*}
\centering
\includegraphics[width=0.9\linewidth]{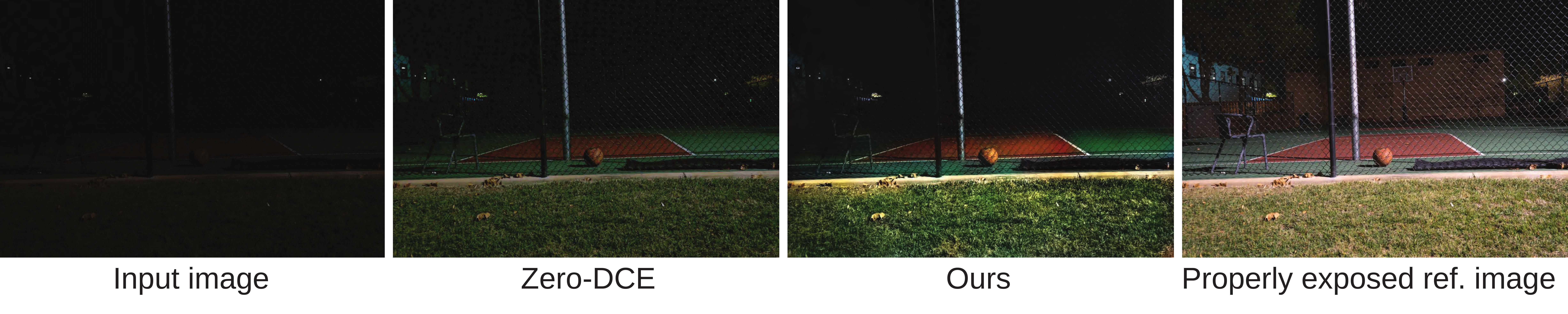}
\vspace{-2mm}
\caption{Qualitative example from the SID dataset \cite{chen2018learning}. We compare our result with the recent Zero-DCE method \cite{guo2020zero}. \vspace{-2mm}}
\label{fig:sid}
\end{figure*}

Fig.\ \ref{fig:qualitative_comparisons_on_DICM} shows a qualitative comparison using the DICM image set. Fig.\ \ref{fig:sid} shows a qualitative comparison the SID dataset \cite{chen2018learning}. In the shown example, we rendered the raw-RGB images provided in the SID dataset to 8-bit JPEG compressed sRGB image. This 8-bit compressed format is more challenging compared to dealing with the 12-bit linear raw images as used by prior work. Though our method is not targeting this kind of ``dark'' scenes, it is arguable that our result is visually on par with the recently proposed method for low-light image enhancement---namely, the Zero-DCE method \cite{guo2020zero}. 

\begin{figure*}[t]
\centering
\includegraphics[width=0.9\linewidth]{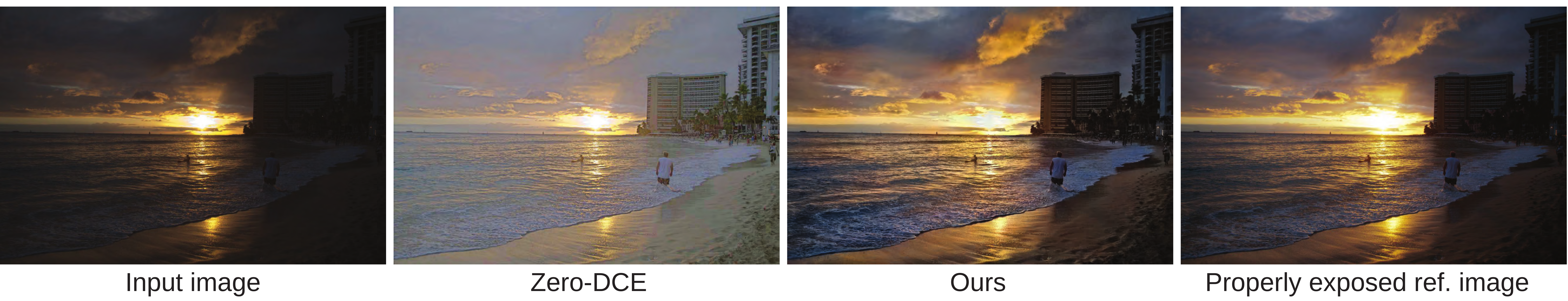}
\vspace{-2mm}
\caption{Qualitative comparison with the recent Zero-DCE method \cite{guo2020zero} on the testing set, used in \cite{DPE}.}
\label{fig:comparisons_with_zero_dce}
\end{figure*}

We further examined our model on the testing set used in \cite{DeepUPE}. This set has no overlap with our training examples taken from the MIT-Adobe FiveK dataset \cite{fivek} and its input images were processed using a different rendering/degradation procedure, as described in \cite{DeepUPE}. Fig.\ \ref{fig:comparisons_with_zero_dce} shows a qualitative comparison between our method and the recent Zero-DCE method \cite{guo2020zero} for low-light image enhancement. The quantitative results using the testing set used in \cite{DeepUPE}, are reported in Table\ \ref{table:new_test_set}. 

\begin{figure*}
\centering
\includegraphics[width=0.9\linewidth]{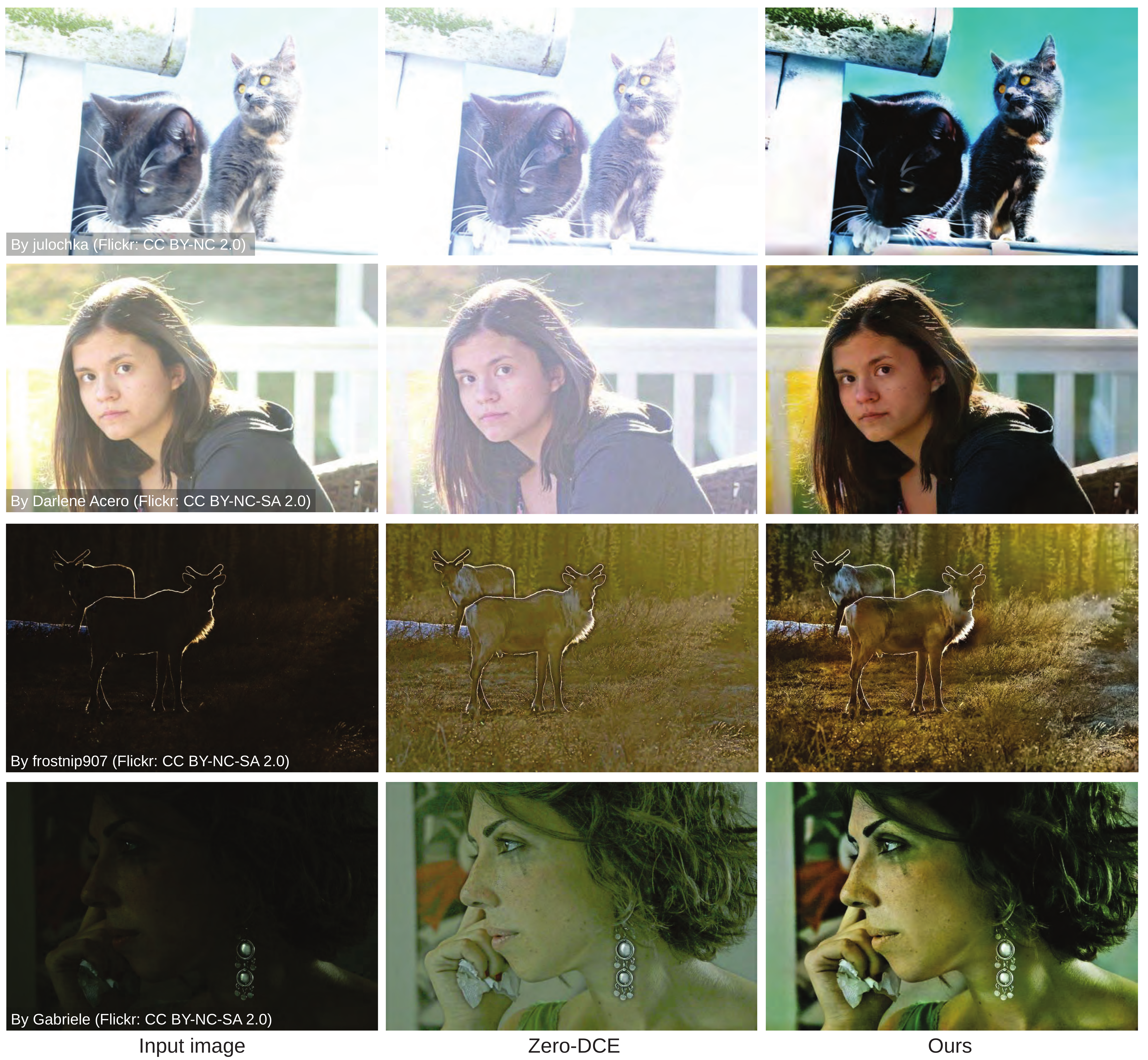}
\vspace{-2mm}
\caption{Comparison with the recent Zero-DCE method \cite{guo2020zero} using images taken from Flickr. \vspace{-2mm}}
\label{fig:flickr}
\end{figure*}

As can be seen, our method achieves on par, sometimes better, results compared to the state-of-the-art methods designed specifically to deal with underexposure errors. Unlike these methods, our method can effectively deal with both under- and overexposure errors, as discussed in the main paper. Note that we did not re-train our method on either the SID dataset or the testing set used in \cite{DeepUPE}, before reporting our results.  Additional qualitative comparisons using images taken from Flickr are shown in Fig.\ \ref{fig:flickr}.

\begin{table}[]
\caption{Comparison with other methods for low-light image enhancement using the test set used in \cite{DeepUPE}. \label{table:new_test_set}}.
\centering
\scalebox{0.84}{
\begin{tabular}{|c|c|}
\hline
Method & PSNR \\ \hline
White-Box \cite{hu2018exposure} & 18.57 \\ \hline
Distort-and-Recover \cite{park2018distort} & 20.97 \\ \hline
Deep UPE \cite{DeepUPE} & \cellcolor[HTML]{79CC7A}\textbf{23.04} \\ \hline
Zero-DCE \cite{guo2020zero}  & 15.455 \\ \hline
Ours &  21.02 \\ \hline
\end{tabular}}
\end{table}

\subsection{Potential Applications}\label{sec:applications}

\begin{figure*}
\centering
\includegraphics[width=0.9\linewidth]{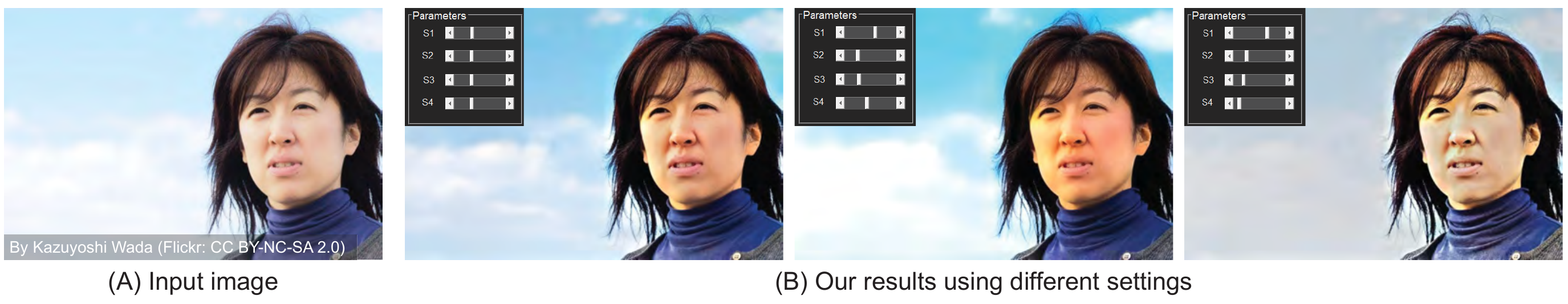}
\vspace{-2mm}
\caption{Our GUI photo editing tool. (A) Input image. (B) Our results using different pyramid level scaling settings set by the user in an interactive way. The input image is taken from Flickr.\vspace{-2mm}}
\label{fig:gui}
\end{figure*}

\begin{figure*}
\centering
\includegraphics[width=0.95\linewidth]{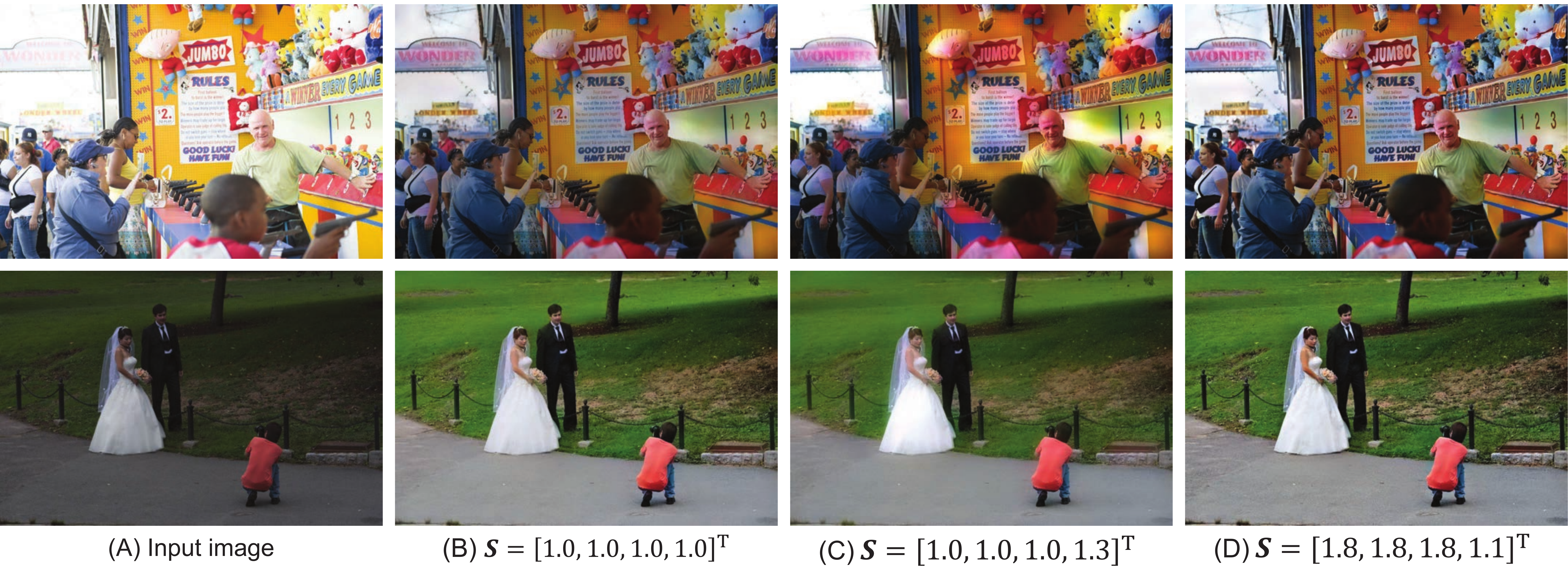}
\vspace{-2mm}
\caption{The effect of the scale vector $\mathbf{S}$ on our final results. (A) Input images. (B-D) Our results using different scale values, $\mathbf{S}$. The shown input images are taken from our validation set.\vspace{-2mm}}
\label{fig:hyperparams}
\end{figure*}

 \begin{figure*}
 \centering
 \includegraphics[width=0.9\linewidth]{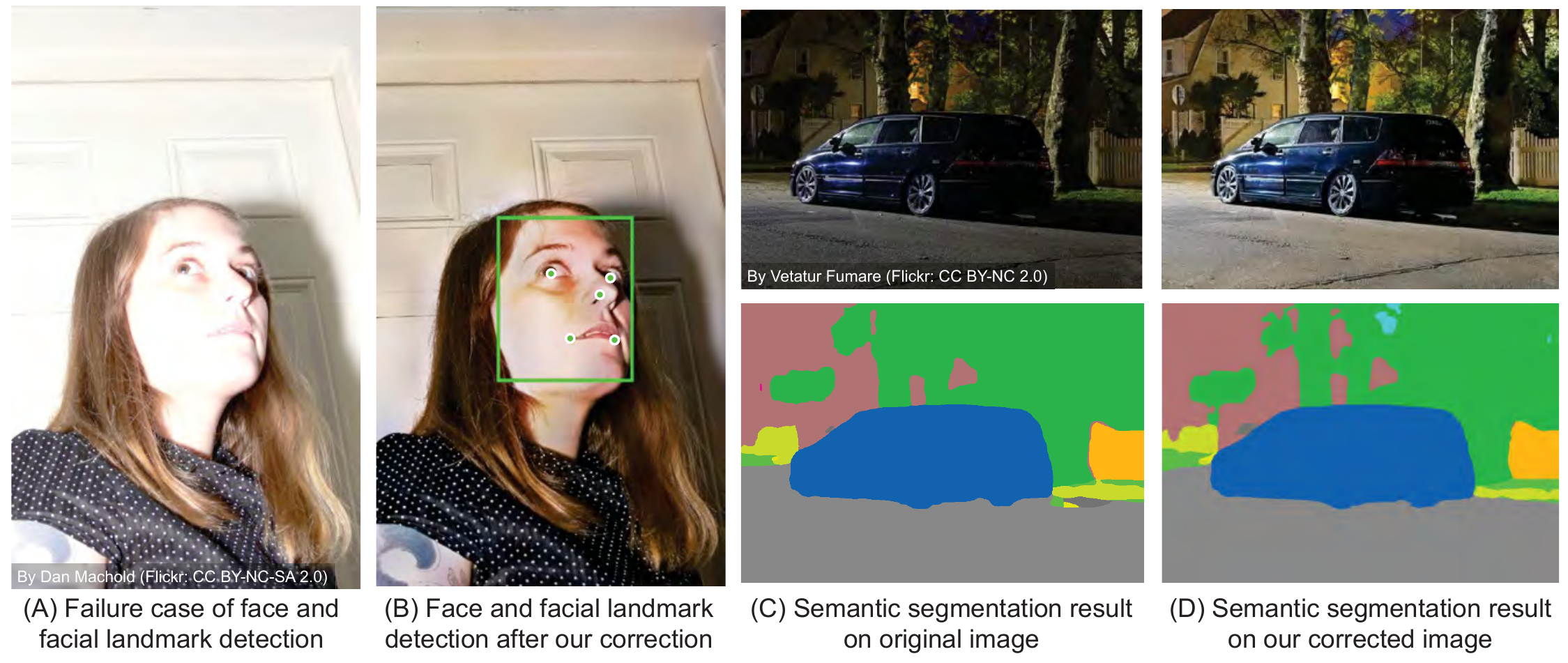}
 \vspace{-2mm}
 \caption{Applying our method as a pre-processing step can improve results of different computer vision tasks. (A) False negative result of face and facial landmark detection due to the overexposure error in the input image. (B) Our corrected image and the results of face and facial landmark detection. (C) Underexposed input image and its semantic segmentation mask. (D) Our corrected image and its semantic segmentation mask. We use the cascaded convolutional networks proposed in \cite{zhang2016joint} for face and facial landmark detection. For image semantic segmentation, we use RefineNet \cite{Lin:2017:RefineNet, lin2019refinenet}. The input images are taken from Flickr.\vspace{-2mm}}
 \label{fig:applications}
 \end{figure*}
 
In this section, we highlight two potential applications of our method: (i) photo editing and (ii) image preprocessing.
 
\paragraph{Photo Editing}
The main potential application of the proposed method is to post-capture correct exposure errors in images. This correction process can be performed in a fully automated way (as described in the main paper) or can be performed in an interactive way with the user. Specifically, we introduce a scale vector $\mathbf{S} = \left[S_1, S_2, S_3, S_4\right]^\top$ that can be used to independently scale each level in the pyramid $\mathbf{X}$ in the inference stage. The scale vector $\mathbf{S}$ is introduced to produce different visual effects in the final result $\mathbf{Y}$. In particular, this scaling operation is performed as a pre-processing of each level in the pyramid $\mathbf{X}$ as follows: $\mathbf{S}_{(l=i)}\mathbf{X}_{(l=i)},$ s.t. $i \in \left\{1, 2, 3, 4\right\}$. 
The values of the scale vector $\mathbf{S}$ can be interactively controlled by the user to edit our network results. Fig.\ \ref{fig:gui} shows different results obtained by our network in an interactive way through our graphical user interface (GUI). Our GUI can be used as a photo editing tool to apply different visual effects and filters on the input images. Note that we used $\mathbf{S} = \left[1.8, 1.8, 1.8, 1.12\right]^\top$ in our experiments in the main paper, as we found it gives the best compelling results (see Fig. \ref{fig:hyperparams}). 

\paragraph{Image Preprocessing}
Our method can also improve the results of computer vision tasks by using it as a pre-processing step to correct exposure errors in input images. Fig.\ \ref{fig:applications} shows example applications. In these examples, we show results of face and facial landmark detection of the work in \cite{zhang2016joint} and image semantic segmentation results obtained by the work in \cite{Lin:2017:RefineNet, lin2019refinenet}. As shown, the results of face detection and semantic segmentation are improved by pre-processing the input images using our method. In future work, we plan to investigate the impact of our exposure correction method on a variety of computer vision tasks.

\end{document}